\documentclass[times,final,twocolumn]{elsarticle}

\usepackage{medima}
\usepackage{framed,multirow}
\usepackage{makecell}

\usepackage{amssymb}
\usepackage{latexsym}
\usepackage[labelfont=bf]{caption}
\usepackage{subcaption}

\usepackage{url}
\usepackage{xcolor}
\usepackage{svg}
\usepackage{amsmath}

\usepackage{hyperref}
\hypersetup{
    colorlinks=true,
    linkcolor=blue,
}

\usepackage{graphicx}
\usepackage{tabularx,booktabs}

\usepackage{amsmath,amssymb,amsfonts}
\usepackage[T1]{fontenc}

\usepackage[table, dvipsnames]{xcolor}

\usepackage{graphicx}
\usepackage{amssymb}
\usepackage{booktabs}
\usepackage{multirow}
\usepackage{setspace}
\usepackage{pdflscape}
\usepackage{rotating}

\usepackage{hyphenat}

\usepackage{colortbl}

\usepackage{cleveref}

\definecolor{C1}{HTML}{FDF1B8} 
\definecolor{C2}{HTML}{E1E1E1} 
\definecolor{C3}{HTML}{F5D5B8} 

\newcommand{\colorboxlegend}[2]{\textcolor{#1}{\rule{#2}{#2}}}

\journal{Medical Image Analysis}

\begin{document}

\verso{Bailiang Jian \textit{et~al.}}

\begin{frontmatter}

\title{
Disentangling Progress in Medical Image Registration: Beyond Trend-Driven Architectures towards Domain-Specific Strategies
}

\author[1,2]{Bailiang~\snm{Jian}\textsuperscript{$\heartsuit$}}
\author[1]{Jiazhen~\snm{Pan}\textsuperscript{$\clubsuit$}}
\author[4]{Rohit~\snm{Jena}}
\author[1,2]{Morteza \snm{Ghahremani}}
\author[1,5]{Hongwei~Bran~\snm{Li}}
\author[1,2,3]{Daniel~\snm{Rueckert}}
\author[1,2]{Christian~\snm{Wachinger}\textsuperscript{$\spadesuit$}}
\author[1,2]{Benedikt \snm{Wiestler}\textsuperscript{$\spadesuit$}}

\nonumnote{$\spadesuit$ Equal Advising.}
\nonumnote{$\clubsuit$ Corresponding Author.}
\nonumnote{$\heartsuit$ Author Contact: \textit{\{bailiang.jian,\ jiazhen.pan\}@tum.de}}

\address[1]{Technical University of Munich, Munich, Germany}
\address[2]{Munich Center for Machine Learning (MCML), Munich, Germany}
\address[3]{Imperial College London, London, England}
\address[4]{University of Pennsylvania, Pennsylvania, USA}
\address[5]{National University of Singapore, Singapore}

\begin{abstract}
Medical image registration drives quantitative analysis across organs, modalities, and patient populations. Recent deep learning methods often combine low-level ``trend-driven'' computational blocks from computer vision, such as large-kernel CNNs, Transformers, and state-space models, with high-level registration-specific designs like motion pyramids, correlation layers, and iterative refinement. Yet, their relative contributions remain unclear and entangled. This raises a central question: should future advances in registration focus on importing generic architectural trends or on refining domain-specific design principles?
Through a modular framework spanning brain, lung, cardiac, and abdominal registration, we systematically disentangle the influence of these two paradigms. Our evaluation reveals that low-level ``trend-driven'' computational blocks offer only marginal or inconsistent gains, while high-level registration-specific designs consistently deliver more accurate, smoother, and more robust deformations. These domain priors significantly elevate the performance of a standard U-Net baseline, far more than variants incorporating ``trend-driven'' blocks, achieving an average relative improvement of $\sim3\%$.
All models and experiments are released within a transparent, modular benchmark that enables plug-and-play comparison for new architectures and registration tasks~(\url{https://github.com/BailiangJ/rethink-reg}). This dynamic and extensible platform establishes a common ground for reproducible and fair evaluation, inviting the community to isolate genuine methodological contributions from domain priors. Our findings advocate a shift in research emphasis: from following architectural trends to embracing domain-specific design principles as the true drivers of progress in learning-based medical image registration.
\end{abstract}


\end{frontmatter}
\vfill
\noindent\textit{Preprint under review at Medical Image Analysis}

\clearpage
\section{Introduction}\label{sec:intro}
\begin{figure}[ht!]
    \centering
    \includegraphics[width=1.0\linewidth]{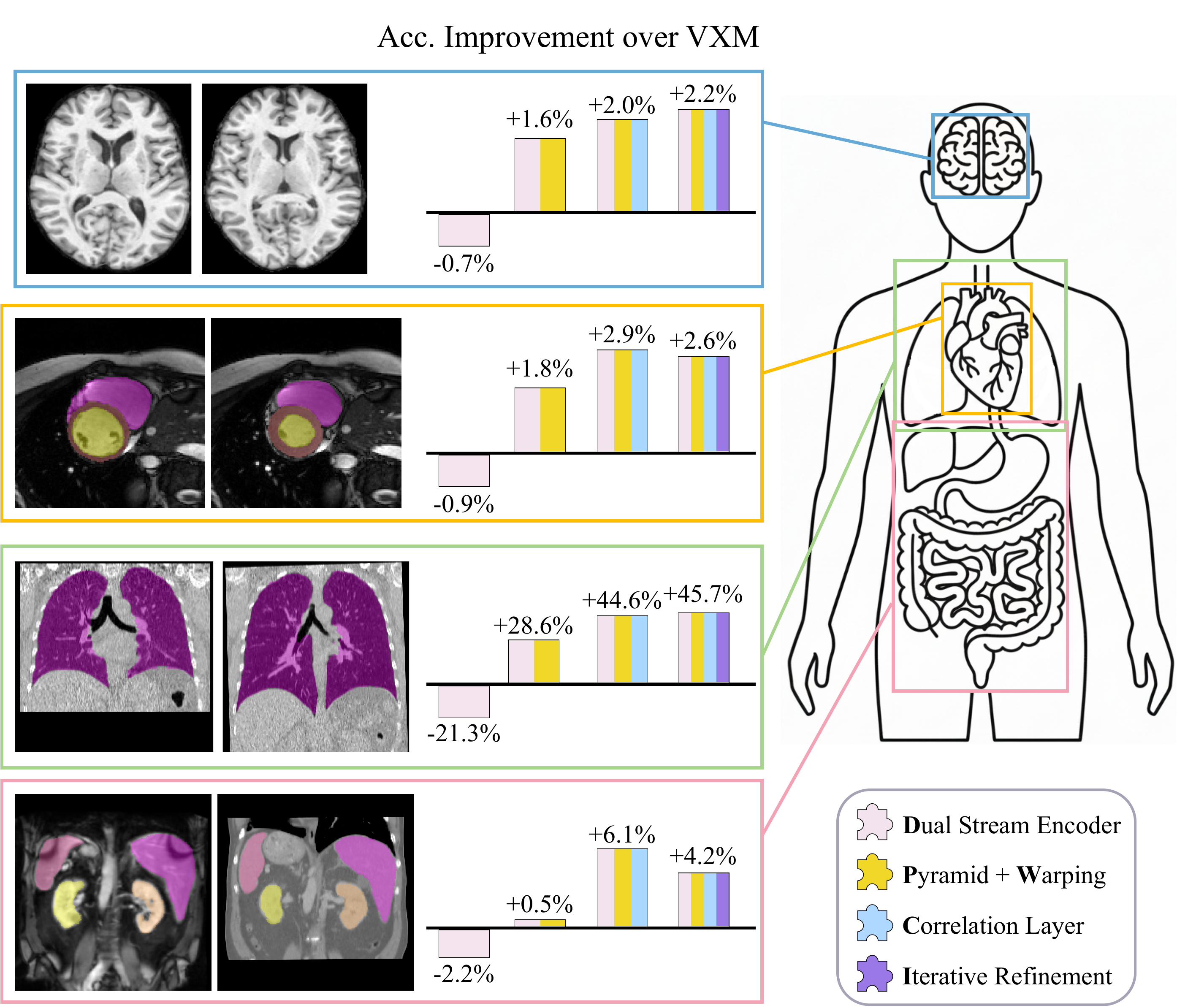}
    \caption{Registration-specific designs substantially enhance registration performance (up to \textbf{45.7\%}) over the VoxelMorph (VXM) baseline across a wide range of registration benchmarks covering multiple modalities and anatomies.
    }
    \label{fig:fig1}
\end{figure}
Medical image registration drives numerous clinical and research applications by estimating spatial correspondences between scans from different timepoints, subjects, modalities, or acquisitions~\citep{jian2025timeflow,avants2008syn,pluim2003mutualinfo}. Accurate registration enables quantitative analysis, disease monitoring, and treatment planning~\citep{im2025longradio,baheti2021bratsreg,brock2017reginradiotherapy} across anatomies ranging from the brain and lungs to the heart and abdomen~\citep{risholm2011mmneuro,reinhardt2008lungreg,makela2002cardiacreg,xu2016abdomen}. With the emergence of deep learning, deformable registration has shifted from hand-crafted similarity metrics and iterative optimization to data-driven networks capable of learning spatial transformations directly from image pairs~\citep{balakrishnan2019voxelmorph}.

Despite impressive progress, a central question remains: what truly drives progress in learning-based medical image registration? Recent research has pursued mainly two paths:

The first emphasizes the adoption of ``trend-driven'' computational blocks: modules imported from the broader computer vision community, such as Vision Transformers~\citep{dosovitskiy2020vit}, large-kernel CNNs~\citep{ding2022replk}, Vision MLPs~\citep{tolstikhin2021mlp-mixer}, and more recently, State Space Models (Mamba)~\citep{gu2024mamba}. These blocks are introduced to capture long-range dependencies, expand receptive fields, mix spatial tokens efficiently, or perform linear-time sequence modeling. However, these general-purpose blocks were not developed specifically for the registration task.

The second revisits the incorporation of registration-specific designs, often well-established in ``traditional'', optimization-based registration: principles developed decades ago to address the geometric nature of image alignment. These designs include dual-stream encoding to project images into a shared latent space for matching, correlation layers to compute dense voxel-wise similarities, multi-resolution pyramids to handle large displacements, and iterative refinement to progressively improve deformation fields~\citep{rueckert1999nonrigid,brox2004high,hosni2012fast}.

In practice, most modern architectures combine both directions, resulting in intertwined contributions, raising the question: are performance gains driven primarily by novel computational blocks, or by classic, well-established registration-oriented strategies? A glimpse of two related domains offers compelling evidence. In medical image segmentation, nnU-Net demonstrated that well-optimized CNN-based U-Nets consistently outperform Transformer- or Mamba-based variants across diverse benchmarks~\citep{isensee2021nnunet,isensee2024nnunetrevisited}, underscoring the power of task-aligned design over architectural novelty. Likewise, in natural language processing, state-of-the-art large language models (LLMs) have advanced through scaling in parameter size and data size, while sharing the same domain-grounded core principle: attention mechanism in Transformers~\citep{vaswani2017attention,raschka2025bigllm}. This parallel motivates our central hypothesis: progress in medical image registration may depend more on registration-oriented priors than on importing the latest architectural trends.

To investigate this hypothesis systematically, we develop a modular framework that disentangles and quantifies the respective contributions of ``\textit{trend-driven}'' and \textit{registration-specific} components. We evaluate methods across four representative registration tasks, each posing a unique challenge~(\Cref{fig:fig1}): (i) subtle anatomical alignment in brain MRI registration, (ii) large displacement, sliding-boundaries in inhale-exhale lung CT registration, (iii) dynamic spatio-temporal deformation in cardiac MRI registration, and (iv) large inter-organ shifts and intensity mismatches in abdominal MRI-CT registration. This task diversity, together with both in-domain and out-of-domain (zero-shot) evaluations, enables a holistic examination of model behavior across deformation scales, modalities, and clinical contexts.
To ensure maximal fairness and transparency, all experiments follow a unified unsupervised training setup with identical preprocessing and loss functions. This extends our previous work~\citep{jian2024rethinkreg}, where weak supervision via Dice loss on brain labels was used. While such supervision can encourage anatomically faithful alignment within a dataset, \cite{jena2024mirage,liu2025unsuprevisit} have shown that it does not translate to robustness under domain shift, as networks may overfit to dataset-specific label distribution. Furthermore, we expand beyond single-study evaluation by introducing an open, dynamic benchmark that hosts all models, configurations, training and evaluation processes. The benchmark enables plug-and-play integration of new methods and tasks, fostering reproducible, transparent, and fair comparisons across both in-domain and zero-shot generalization settings.

Our key contributions are threefold:
\begin{itemize}
    \item \textbf{Modularized Disentanglement}: A modular framework that isolates and systematically evaluates the impact of low-level ``\textit{trend-driven}'' computation blocks versus high-level \textit{registration-specific} designs, enabling fine-grained performance attribution.
    \item \textbf{Cross-Domain Empirical Analysis}: Extensive experiments across brain, lung, cardiac, and abdomen, covering mono-modal and multi-modal settings, as well as both in-domain and out-of-domain (zero-shot) testing, reveal that domain-specific designs are the consistent drivers of accuracy, smoothness, and generalizability.
    \item \textbf{Open, Extensible Benchmark}: A public and extensible benchmark with all code and results released at \url{https://github.com/BailiangJ/rethink-reg}, establishing a foundation for fair, reproducible, and continuously evolving registration research.
\end{itemize}

\section{Related Work}\label{sec:rel_work}
\subsection{Trend-driven Blocks Imported from Computer Vision}
Architectural innovations from computer vision have been increasingly adopted for deformable registration. These blocks typically aim to capture long-range dependencies, enlarge receptive fields, or improve sequence modeling efficiency, often by replacing the convolutional encoder in U-Net-style architectures.

\subsubsection{Transformers and MLP-based Models}
The Vision Transformer (ViT)~\citep{dosovitskiy2020vit} represents images as sequences of patches and models global context via self-attention. Early studies~\citep{chen2021vit-v-net,mok2022c2fvit} incorporated ViT for medical image registration, motivated by its ability to capture long-range dependencies and address large displacements. However, the quadratic computational cost of global attention limits the direct use of ViT in volumetric settings. To mitigate this, the Swin Transformer~\citep{liu2021swin}, which computes attention within shifted, localized windows, has been widely adopted. \cite{chen2022transmorph} proposed TransMorph, where the Swin Transformer encoder replaces the convolutional encoder, improving the volumetric registration capacity. Later extensions introduced cross-attention between source and target features or across multiple scales~\citep{song2022attenreg,shi2022xmorpher,chen2023transmatch,ghahremani2024hvit}. Beyond attention-based models, \cite{tolstikhin2021mlp-mixer} proposed MLP-Mixer for spatial and channel mixing of image patches, reaching effective global context as well. Building on this, \cite{meng2024corr-wmlp} adapted MLPs to capture global dependencies in registration. While promising, these methods' architectural designs often incorporate registration-specific components like multi-resolution pyramids or correlation layers, making it difficult to isolate the precise contribution of the attention or MLP mechanism alone.

\subsubsection{Convolutional Variants}
VoxelMorph~\citep{balakrishnan2019voxelmorph} established the convolutional U-Net~\citep{ronneberger2015u} as a widely used baseline for learning-based registration. Building on this, researchers have explored variations beyond the standard $3\times 3$ convolutional operator. \cite{miao2018dilatedconvreg} employed dilated convolutions~\citep{yu2015dilatedconv} to expand the receptive field without increasing parameters in registration networks, while \cite{jia2022lku} used large-kernel convolutions~\citep{ding2022replk} to approximate long-range dependencies similar to Transformer-based networks. Moreover, \cite{huang2024deformableconvreg} utilized deformable convolutions~\citep{dai2017deformableconv} to allow adaptive kernel shapes for more flexible feature extraction during registration, while \cite{cheng2025sacb,zheng2024ran} adjusted kernel weights to adapt to anatomical or modality-specific variations. While these approaches enhance the convolutional paradigm, they raise the question of how these incremental improvements to the backbone compare to the impact of higher-level architectural designs.

\subsubsection{State-Space Models (Mamba)}
Mamba~\citep{gu2024mamba} is a selective state-space model (SSM) that enables dynamic selectivity in sequential modeling. Its hardware-aware algorithm allows nearly linear complexity with respect to sequence length. \cite{guo2024mambamorph} introduced MambaMorph by substituting CNNs with Mamba blocks in the VoxelMorph's encoder, and subsequent works~\citep{hu2024regmamba,wen2024mambareg,chen2024lungmambaconvnet,chen2025mambabir,xu2025hybridmorph,xue2025mambasubstrate} explore Mamba for brain MRI, lung CT, and Abdomen CT registration. These studies posit that the efficient modeling of long-range dependencies by Mamba improves the registration results. However, the benefits usually come together with registration-specific designs like image pyramid, leaving the true benefit of Mamba for registration as an open question. A recent revisiting study on nnU-Net~\citep{isensee2024nnunetrevisited} reported that replacing convolution with Mamba layers~\citep{ma2024umamba} did not affect segmentation performance. The reported gains primarily stemmed from residual connections in the U-Net encoder.

\subsubsection{Other Imported Paradigms}
Beyond backbones, several other advanced techniques have been adapted for registration. \cite{wolterink2022inrreg,van2023inrcycle,van2024reginr,sideri2024sinr} adopted Implicit Neural Representations (INRs)~\citep{sitzmann2020inr} to parametrize flow fields with coordinate-based MLPs, while \cite{xu2021msode,wu2022nodeo} add NeuralODE~\citep{chen2018neuralode} on top to integrate the velocity field and achieve smoother and more accurate deformations. Besides, \cite{kim2022diffusemorph, qin2023fsdiffreg} exploit the latent features learned in the Diffusion-based generative models to further refine the registration. Together, these methods illustrate the breadth of paradigms imported from computer vision into registration, even though their specific contributions are often intertwined with domain-specific adaptations.

\subsection{Registration-specific Architectural Design}
In parallel to trend-driven imports, registration-specific designs explicitly encode domain knowledge about the image alignment task itself. These strategies build primarily on principles established in classical registration and optical flow estimation methods, and have been widely inherited and extended in deep learning frameworks.

\subsubsection{Dual-stream/Siamese Encoders}
Instead of concatenating the source and target images at the input, dual-stream encoders process them independently with shared or separate weights. This design is common in optical flow networks~\citep{sun2018pwc,teed2020raft} and has been adapted in registration frameworks such as \cite{pan2021efficient,meng2022nice,shi2022xmorpher,pan2024unrolled}. It encourages the extraction of feature representations that remain discriminative within each image but comparable across images, thereby facilitating explicit correspondence matching, as in traditional feature-based image matching methods~\citep{lowe2004sift,bay2006surf}.

\subsubsection{Motion Pyramid and Warping}
Coarse-to-fine refinement is a classical paradigm in registration~\citep{rueckert1999nonrigid,avants2008syn,vercauteren2009demons}. Deformable image registration is inherently a high-dimensional, nonlinear, and non-convex optimization problem, often leading to ill-conditioning and susceptibility to local minima. Multi-resolution optimization mitigates these challenges by gradually increasing spatial resolution, analogous to simulated annealing or scale-space optimization, which improves convergence to better optima~\citep{jena2024fireants}. Deep learning counterparts adopt hierarchical pyramids of images or features: deformations are first estimated at coarse resolution to capture global motion, and then progressively refined at finer scales using warped features or images from the previous stage~\citep{mok2020lapirn,kang2022prnet,lv2022joint,meng2022nice,yin2023pc}. This design is particularly effective for capturing large displacements.

\subsubsection{Correlation Layers / Cost Volumes}
Correlation volumes, long established in optical flow estimation~\citep{brox2004high,hosni2012fast,dosovitskiy2015flownet,sun2018pwc,teed2020raft}, have been adapted for solving voxel-wise correspondence in registration. \cite{heinrich2019pddnet} approximated cost volumes via localized MSE, while later works~\citep{wang2023modet,liu2024vfa,jian2024rethinkreg} directly compute inner products between features within search windows. Correlation is often paired with dual-stream encoders, enabling explicit voxel-wise comparison between target and source features rather than relying on implicit fusion.

\subsubsection{Iterative/Recursive Refinement}
Complementing the multi-scale approach, iterative refinement improves alignment by applying recursive updates at a fixed resolution. Inspired by recurrent updates in optical flow~\citep{teed2020raft}, this strategy iteratively refines the deformation by repeatedly feeding the currently warped image or features back into a refinement network or a decoder block to predict incremental updates~\citep{zhao2019recursive,jia2021vrnet,ma2023pivit,wang2024rdp,ma2024iirp}.  This process progressively corrects residual misalignment at the same scale, and conceptually approximates velocity integration techniques such as scaling-and-squaring, often yielding smoother and more accurate deformations than a single, one-shot prediction.

\subsubsection{Other Task-specific Strategies}
Other strategies aim to enforce deformation plausibility and consistency. Symmetric registration~\citep{avants2008syn,mok2020symnet} jointly deforms both images to a mid-space, while inverse consistency can be enforced either by construction~\citep{greer2023byconstructicon,honkamaa2023sitreg} or by extra constraints~\citep{kim2021cyclemorph,greer2021icon,van2023inrcycle,tian2023gradicon,reithmeir2025regularizationreview}. Parametric formulations based on splines~\citep{qiu2021vxmbspline,wang2023keymorph} or Gaussian primitives~\citep{li2025gaussian,tian2025gaussian} further provide smooth, compact, and efficient parameterizations of the deformation field. Collectively, these strategies illustrate the diversity of task-specific solutions developed to improve alignment quality and regularity.

\subsection{Registration Benchmarks}
Public registration challenges have played an essential role in advancing the field by providing standardized datasets, tasks, and evaluation protocols across anatomies and modalities. Notable examples, including EMPIRE10~\citep{murphy2011empire10}, Learn2Reg~\citep{hering2022learn2reg,hansen2025learn2reg2024}, ReMIND2Reg~\citep{dorent2025remind2reg}, and LUMIR~\citep{chen2025lumir} cover anatomies including brain, lung, chest, and abdomen, and modalities including MRI, CT, and Ultrasound. These benchmarks offer a clear snapshot of the state-of-the-art on difficult clinical problems and have fostered rapid methodological innovation. 
At the same time, the primary goal of challenge participation is to maximize leaderboard performance. To achieve this, competing solutions often integrate a range of architectural choices, supervision schemes, and training strategies into complex pipelines. While effective, this makes it difficult to attribute performance gains to specific design components, obscuring the individual contributions of trend-driven modules or registration-specific elements. 
This observation motivates our study: rather than evaluating holistic solutions, we propose a modular framework that isolates and quantifies the effects of different architectural components under identical experimental setups. In doing so, we aim to provide a clearer understanding of which design choices truly drive performance in deformable registration.

\section{Methodology}\label{sec:method}
\begin{figure*}[ht!]
    \centering
    \includegraphics[width=1.0\linewidth]{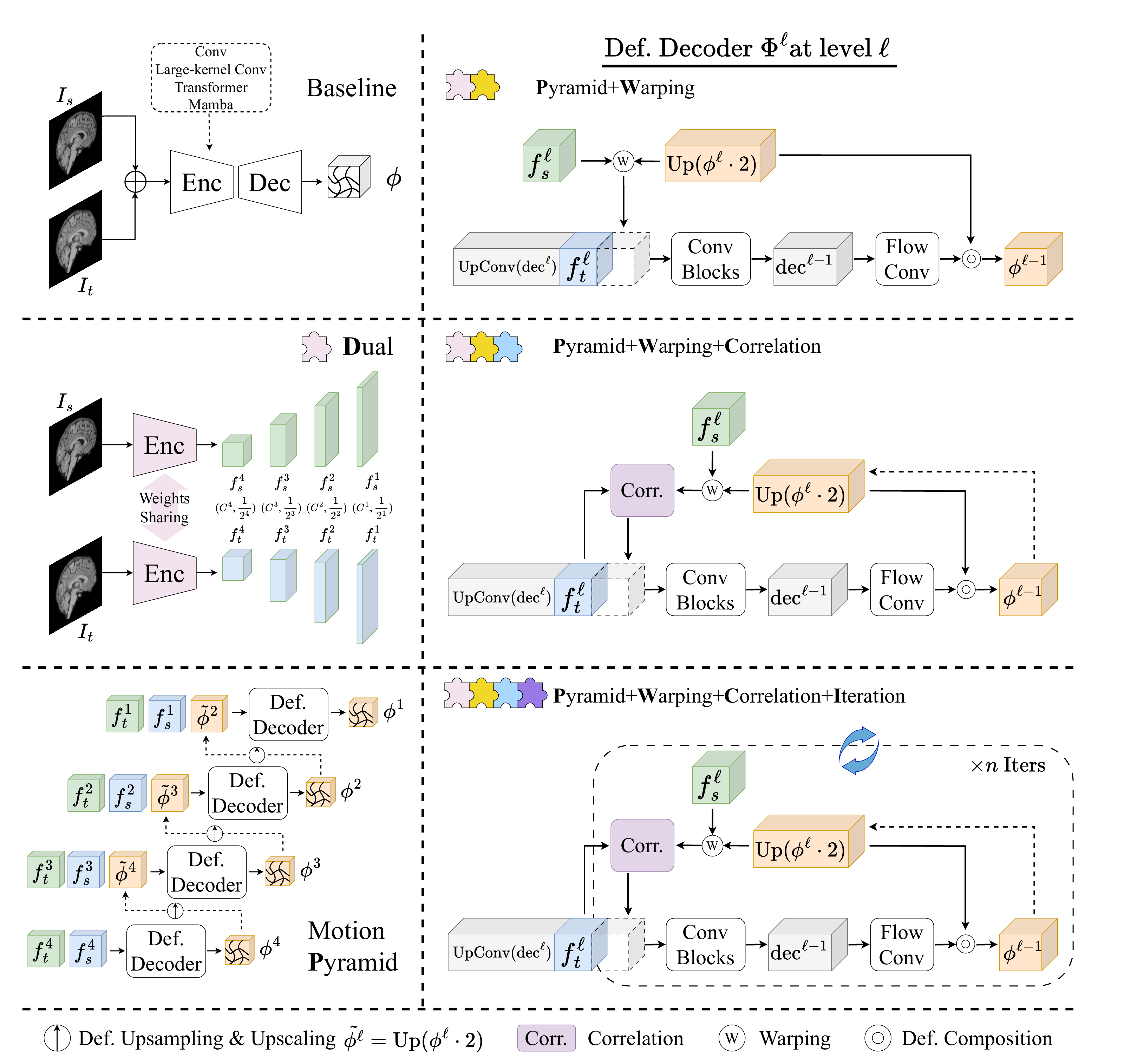}
    \caption{
    Overview of the baseline and the modular registration components. \textbf{Upper Left}: the \textit{Baseline} architecture concatenates image pairs and predicts the deformation field directly. \textbf{Middle Left}: \textit{\textbf{D}ual}-stream encoder extracts multi-resolution feature pyramids for the source and target images separately. \textbf{Bottom Left}: \textit{Motion \textbf{P}yramid} refines the deformation field progressively from coarse to fine using hierarchical features. \textbf{Right}: workflow of the deformation decoder at level $\ell$. Level $\ell$ corresponds to $2^{-\ell}$ resolution. Registration-specific designs are cumulatively integrated: \textbf{Top}: \textit{\textbf{P}yramid and \textbf{W}arping}; \textbf{Middle}: \textit{\textbf{C}orrelation}; \textbf{Bottom}: \textit{\textbf{I}teration}. The cube widths are illustrative and do not represent the exact channel dimensions.
    }
    \label{fig:method}
\end{figure*}

\subsection{Preliminaries}
Let $I_t, I_s: \Omega \to \mathbb{R}$ be the target (or fixed) and source (or moving) images, respectively, defined over a spatial domain $\Omega \subset \mathbb{R}^d$, where the dimension $d\in\mathbb{N}$ is typically 2 or 3. Deformable image registration aims to find a dense deformation field $\phi: \Omega \to \Omega$ that establishes a spatial correspondence between the two images.

In a deep learning-based framework, this is achieved by training a neural network $\Phi$, parametrized by weights $\theta$, to predict the deformation field from the input image pair:
\begin{equation}
    \phi = \Phi_\theta(I_t, I_s).
\end{equation}
In this work, we adopt the common backward (or pullback or Euler coordinates) convention: $\phi$ maps each coordinate $\mathbf{x} \in \Omega$ in the target image's space to its corresponding coordinate in the source image's space. The source image is then warped according to this field to align with the target. The warped source (or moved) image $I_m$ is defined as:
\begin{equation}
    I_m(\mathbf{x})=(I_s \circ \phi)(\mathbf{x}) := I_s(\phi(\mathbf{x})).
\end{equation}
Equivalently, the deformation field can be parametrized via an additive displacement field $\mathbf{u}:\Omega \to \mathbb{R}^d$, $\phi(\mathbf{x})=\mathbf{x}+\mathbf{u}(\mathbf{x})$.
The objective is to find a $\phi$ such that the moved image $I_m$ is similar to the target $I_t$, i.e., $I_t \approx I_s \circ \phi$.

To optimize the network parameters $\theta$, a loss function $\mathcal{L}$ is minimized via gradient descent. This loss typically consists of two main components:
\begin{equation}
    \mathcal{L}(\phi, I_t, I_s) = \mathcal{L}_\text{sim}(I_t, I_s \circ \phi) + \lambda \mathcal{L}_\text{reg}(\phi).
\end{equation}

The first term, $\mathcal{L}_\text{sim}$, quantifies the dissimilarity between the target image and the warped source (moved) image. Its choice depends on the imaging modalities; common examples include Mean Squared Error (MSE) or Localized Normalized Cross-Correlation (LNCC) for mono-modal registration, and Mutual Information (MI) or MIND-SSC for multi-modal cases.

The second term, $\mathcal{L}_\text{reg}$, is a regularization penalty imposed on the deformation field $\phi$ or the displacement field $\mathbf{u}$. This term ensures the transformation is smooth and physically plausible, penalizes unrealistic distortions such as tears or foldings. Common regularizers are Diffusion (first-order smoothness), Bending energy (second-order smoothness), and a direct penalty on non-positive Jacobian determinants. The hyperparameter $\lambda$ controls the trade-off between image similarity and deformation regularity.

While additional supervision can be incorporated using annotations like segmentation labels, this work focuses on unsupervised training, where the optimization is driven purely by information from the image intensities.

\subsection{Baseline}
Our reference model is the canonical U-Net-based architecture introduced by VoxelMorph~\citep{balakrishnan2019voxelmorph}, a fully-convolutional encoder-decoder network with skip connections.
Conceptually, this architecture operates as a single-stream model where the target and source images, $I_t$ and $I_s$, are concatenated and fed into the network as a single tensor. The encoder progressively extracts hierarchical features using a series of standard 3$\times$3 convolutional blocks, and the decoder upsamples and refines these features to predict a final voxel-wise displacement field $\mathbf{u}$. This displacement defines the dense deformation field via $\phi(\mathbf{x})=\mathbf{x}+\mathbf{u}(\mathbf{x})$, which is then used to warp the source image.

This simple yet powerful CNN-based U-Net serves as our baseline framework. It enables systematic integration of both trend-driven computational blocks (e.g., Vision Transformers, Mamba) and registration-specific designs (e.g., dual-stream image encoder, motion pyramid, correlation layers). All subsequent variants are derived by modular substitution or extension of this baseline.

\subsection{Modular Framework}
To disentangle the contributions of computational novelty versus registration-specific design, we adopt a modular evaluation strategy. Each candidate block is integrated into the baseline U-Net in isolation or in controlled combinations, while all other components (loss functions, training datasets, hyperparameters, and random seeds) are held fixed.

Formally, let $\mathcal{M}_\text{base}$ denote the baseline U-Net registration model (VoxelMorph). A trend-driven block $\mathcal{B}_\text{trend}$ or a registration-specific block $\mathcal{B}_\text{reg}$ can be substituted into the baseline to yield a new variant:
\begin{equation*}
    \mathcal{M}=\mathcal{M}_\text{base} \oplus \mathcal{B}_1 \oplus \mathcal{B}_2 \oplus \cdots \oplus \mathcal{B}_n,
\end{equation*}
where each $\mathcal{B}_i \in \{\mathcal{B}_\text{trend}, \mathcal{B}_\text{reg}\}$ represents a specific architectural component, and $\oplus$ denotes modular replacement or augmentation. By progressively stacking blocks from either category, we can measure both marginal contributions (the effect of each block individually) and interaction effects (when multiple blocks are combined).

This modular logic yields a factorial comparison design, enabling fair and transparent attribution of performance gains to specific architectural choices. Importantly, every variant is trained and evaluated under identical conditions, ensuring that observed differences reflect only the architectural modifications.

\subsection{Trend-driven Computational Blocks}
Recent advances in computer vision have introduced a series of architectural building blocks designed to expand receptive fields, capture long-range dependencies, or improve sequence modeling efficiency. To assess their impact on deformable registration, we replace the encoder blocks of the baseline U-Net (VoxelMorph) with the following variants.

\paragraph{\textbf{Vision Transformer (TransMorph)}}
To capture global dependencies, TransMorph~\citep{chen2022transmorph} replaces the convolutional encoder with Swin Transformer blocks. The Swin Transformer divides images into non-overlapping patches, similar to the standard Vision Transformer, but restricts self-attention to localized, shifted windows. This design efficiently models both local and global context, while avoiding the quadratic complexity of full attention. The decoder remains convolutional for comparability. This variant is denoted as \textbf{TM}.

\paragraph{\textbf{Large-Kernel Convolution (LKU-Net)}}
Larger convolutional kernels can approximate long-range dependencies~\citep{ding2022replk}. LKU-Net~\citep{jia2022lku} augments each encoder block with a parallel $5\times 5$ convolution branch, effectively enlarging the receptive field while retaining the efficiency of convolutions. This variant is denoted as \textbf{LKU}.

\paragraph{\textbf{Mamba Encoder}}
Mamba~\citep{gu2024mamba} is a state-space model (SSM) with a hardware-optimized selective scanning mechanism for efficient sequence modeling. MambaMorph~\citep{guo2024mambamorph} first introduced it for deformable registration to handle long-range dependencies more efficiently. We evaluate two integration strategies:
\begin{itemize}
    \item \textbf{Mam-VXM}: Replaces the convolutional encoder of VoxelMorph with stacked Mamba blocks, while retaining the original decoder.
    \item \textbf{Mam-TM}: Replaces the Swin Transformer blocks in TransMorph with Mamba blocks. Patch embedding, positional encodings, and MLP layers are retained to preserve the patch-based formulation.
\end{itemize}
These two settings allow us to disentangle the role of Mamba within convolutional and transformer-style encoders.

\subsection{Registration-specific Design Blocks}
Beyond generic trend-driven blocks, several architectural strategies have been proposed specifically for deformable registration. These strategies encode domain knowledge, such as coarse-to-fine refinement, explicit feature correspondences via correlation, or iterative updates, and are well-established in ``traditional'' registration. We integrate the following four components into our modular framework:

To formalize composition, let $\phi_1,\phi_2: \Omega \to \Omega$ be deformation fields. Their composition is
\begin{equation}
    (\phi_1 \circ \phi_2)(\mathbf{x}) = \phi_2(\phi_1(\mathbf{x})),
\end{equation}
corresponding to applying $\phi_1$ first and then $\phi_2$. In terms of displacement fields $\mathbf{u}_1,\mathbf{u}_2$, this becomes
\begin{equation}
    (\mathbf{u}_1 \circ \mathbf{u}_2)(\mathbf{x}) = \mathbf{u}_1(\mathbf{x}) + \mathbf{u}_2(\mathbf{x} + \mathbf{u}_1(\mathbf{x})).
\end{equation}

\paragraph{\textbf{Dual-Stream Encoders}}
Instead of concatenating the target and source images, a shared-weight encoder $E$ extracts a multi-scale feature pyramid from the source and target images independently: 
\begin{equation}
    \{ f_t^\ell \}_{\ell=0}^L = E(I_t), \quad \{ f_s^\ell \}_{\ell=0}^L = E(I_s).
\end{equation}
where $f_t^\ell, f_s^\ell \in \mathbb{R}^{H_\ell \times W_\ell \times D_\ell \times C_\ell}$ denotes the feature map at resolution level $\ell$, with spatial size $(H_\ell, W_\ell[, D_\ell])$ and channel dimension $C_\ell$, where $D_\ell$ is present only in 3D cases. Level $\ell$ corresponds to scale $2^{-\ell}$ resolution relative to the original image.
This independent encoding encourages discriminative yet comparable features for subsequent matching. Denoted as \textbf{D}.

\paragraph{\textbf{Motion Pyramid and Warping}}
To capture large deformations, we adopt a coarse-to-fine motion pyramid. At the coarsest level $\ell=L$, the decoder predicts $\phi^L$. For each finer level $\ell$, the flow from level $\ell+1$ is upsampled and upscaled:
\begin{equation}
    \tilde{\phi}^{\ell+1} = 2 \cdot \text{Up}(\phi^{\ell+1}).
\end{equation}
The decoder $\Phi^\ell$ then predicts a residual update
\begin{equation}
    \Delta \phi^{\ell} = \Phi^\ell\!\left(f_t^\ell, f_s^\ell \circ \tilde{\phi}^{\ell+1}\right),
\end{equation}
and the flow is refined as
\begin{equation}
    \phi^\ell = \tilde{\phi}^{\ell+1} \circ \Delta \phi^\ell.
\end{equation}
This component is denoted as \textbf{PW}.

\paragraph{\textbf{Correlation Layers}}
To explicitly establish voxel-wise correspondences, we compute local correlations between features at level $\ell$. Let
\begin{equation}
    f_t^\ell, f_s^\ell \in \mathbb{R}^{H_\ell \times W_\ell \times D_\ell \times C_\ell}
\end{equation}
be the target and source feature maps. For each voxel $\mathbf{x}\in\Omega_\ell$, the correlation is defined within a local neighborhood $\mathcal{N}_r(\mathbf{x})=\{\mathbf{y}\in\Omega_\ell : \|\mathbf{y}-\mathbf{x}\|_\infty \le r\}$ of radius $r$ as
\begin{equation}
    C^\ell(\mathbf{x}, \mathbf{y}) = \frac{1}{C_\ell} f_t^\ell(\mathbf{x})^\top f_s^\ell(\mathbf{y}), \quad \mathbf{y}\in \mathcal{N}_r(\mathbf{x}).
\end{equation}
The resulting correlation volume has shape $H_\ell \times W_\ell [\times D_\ell] \times (2r+1)^d$, where $d \in \{2,3\}$ is the spatial dimension. This module is denoted as \textbf{C}.

\paragraph{\textbf{Iterative Refinement}}
Instead of estimating the full deformation field in a single forward pass, we employ iterative refinement at each level. Let $\Phi^\ell$ denote the decoder at level $\ell$. At iteration $k$, given the current estimate $\phi^\ell_{(k)}$, $\Phi^\ell$ predicts an incremental update
\begin{equation}
    \Delta \phi^\ell_{(k)} = \Phi^\ell\!\left(f_t^\ell, f_s^\ell \circ \phi^\ell_{(k)}\right),
\end{equation}
and the field is updated by composition
\begin{equation}
    \phi^\ell_{(k+1)} = \phi^\ell_{(k)} \circ \Delta \phi^\ell_{(k)}.
\end{equation}
Starting from $\phi^\ell_{(0)}=\tilde{\phi}^{\ell+1}=\text{Up}(\phi^{\ell+1})\cdot 2$, this recursive refinement progressively improves alignment. Conceptually, the procedure approximates the integration of an underlying velocity field, yielding smoother deformations. Denoted as \textbf{I}.

\paragraph{\textbf{Modular Integration}}
Each of the above registration-specific components (\textbf{D}, \textbf{PW}, \textbf{C}, \textbf{I}) can be integrated into the baseline independently or in combination. As shown in~\Cref{fig:method}, within our framework, these modules are incrementally added on top of the baseline encoder-decoder, enabling systematic quantification of their individual contributions as well as their synergistic effects.

\section{Experiments}\label{sec:exp}
\subsection{Datasets and Tasks}
We evaluate across four representative deformable registration tasks covering diverse anatomies, modalities, and motion patterns: (1) cross-sectional brain MRI registration, (2) longitudinal lung CT registration, (3) temporal cardiac MRI 2D registration, and (4) intra-subject abdominal MRI–CT registration. This diversity allows us to assess both trend-driven and registration-specific designs under heterogeneous imaging and anatomical conditions.

\subsubsection{Cross-sectional Brain MRI Registration}
This task targets mono-modal brain MRI alignment across different subjects. We use the LUMIR challenge dataset~\cite{chen2025lumir} as the primary training data, taking the first 500 scans for training from 3384 subjects. Evaluation is conducted on the 30 leaderboard validation subjects, which include 10 in-domain, 10 out-of-domain with landmark annotations, and 10 out-of-domain high-field MRI scans with manual segmentations, resulting in 27 validation pairs.  
To test generalization, we further perform zero-shot evaluation on OASIS~\cite{marcus2007oasis} (84 scans), ADNI~\cite{jack2008adni} (43), IXI\footnote{\url{https://brain-development.org/ixi-dataset/}} (115), LPBA~\cite{shattuck2008lpba} (40), and Mindboggle~\cite{klein2005mindboggle} (100) datasets, each with 200 randomly sampled test pairs following~\cite{jian2024rethinkreg}.  
All scans are in 1\,mm isotropic spacing and with image size $160\times192\times224$. Intensities are min–max normalized between the 0th and 99.99th percentiles without clipping.  
Training uses the Local Normalized Cross-Correlation (LNCC) loss with window size 9 and a regularization weight $\lambda=0.5$. The batch size is set to 1, with 100 epochs of 250 iterations each. Left–right flipping ($p=0.5$) is applied as data augmentation.

\subsubsection{Longitudinal Lung CT Registration}
We evaluate intra-patient registration between inspiration and expiration phases using the NLST dataset from the Learn2Reg challenge~(\cite{clark2013tcia,hering2022learn2reg}). The dataset provides paired CT scans, lung masks, and anatomical keypoints. We split the data into 110/50/50 pairs for training, validation, and testing.  
For zero-shot evaluation, we use the Lung250M-4B dataset~(\cite{falta2023lung250m}), consisting of 17 pairs from the validation and test sets (excluding overlap with the NLST dataset) with landmark annotations. All images in the NLST dataset are in 1.5\,mm isotropic spacing and with image size $224\times192\times224$. Because voxel spacing in Lung250M-4B varies, we resize the images to an identical image size $224\times192\times224$. Therefore, we report Target Registration Error (TRE) in millimeters for the NLST dataset and in voxel units for the resampled Lung250M-4B dataset. Intensities are clipped to the Hounsfield Unit (HU) range of [–1000, 500] and normalized to [0, 1]. The LNCC loss (window size = 9) is used with a regularization weight $\lambda=1.0$. Training uses a batch size of 1 for 200 epochs (80 iterations per epoch) and left–right flipping ($p=0.5$) augmentation.

\subsubsection{Temporal Cardiac MRI 2D Registration}
This task evaluates short-axis cardiac motion between end-diastole (ED) and end-systole (ES) frames. We use the ACDC dataset~(\cite{bernard2018acdc}) for training and in-domain testing and the M\&Ms (Multi-Centre, Multi-Vendor \& Multi-Disease) dataset~(\cite{campello2021mandm}) for zero-shot testing. Both datasets include segmentations of the left ventricle (LV), right ventricle (RV), and myocardium (MYO).  
The ACDC dataset is split into 80/20/50 subjects for training, validation, and testing, corresponding to 506/190/194 slices, while the M\&Ms dataset contains 1859 slices from six centers across three countries. Because of thick-slice acquisitions (5–10 mm) along the cardiac long-axis, registration is performed in 2D slices.  
Slices are cropped to the cardiac region of interest based on segmentation masks, resulting in an image size $128\times128$. Intensities are min-max normalized to [0, 1] per slice. The Mean Squared Error (MSE) loss is used with a regularization weight $\lambda=0.05$. Each iteration processes all 2D slices from one 3D volume as a batch. Training runs for 200 epochs with 80 iterations per epoch, with random horizontal and vertical flipping for augmentation.

\subsubsection{Intra-subject Abdomen MRI-CT Registration}
We use the abdominal MRI–CT pairs from the Learn2Reg challenge~(\cite{hering2022learn2reg}), comprising 16 intra-patient cases. Eight pairs include manual annotations of the spleen, liver, and kidneys, which serve as the test set. The remaining eight pairs are segmented using TotalSegmentator~(\cite{wasserthal2023totalsegmentator}), and used for training.  
To remove coarse misalignment and focus on deformable correspondence, we perform an initial rigid translation alignment using organ centroid coordinates. All volumes are in 2 mm isotropic spacing and with image size $192\times160\times192$.  
MRI intensities are min–max normalized between the 0th and 99.99th percentiles, while CT intensities are clipped to the HU range [–450, 450] and normalized to [0, 1].  
The Modality Independent Neighborhood Descriptor (MIND) loss~(\cite{heinrich2013mindssc}) is used with radius = 2, dilation = 2, and a regularization rate $\lambda=1.0$. Training uses batch size 1, 200 epochs (8 iterations per epoch), and left–right flipping ($p=0.5$) as augmentation.

\subsection{Modular Framework and Architectural Configurations}
To ensure a fair comparison, all tested architectures adhere to two universal design principles. First, all models predict final displacement fields at half-resolution, which are then upsampled to the full image grid for warping. Second, all models share a standardized output layer, the \textit{Flow Convolution}: a kernel size 3 convolution with stride 1, output channels equal to the spatial dimension, weight initialization $\mathcal{N}(0,10^{-5})$, and zero bias.

\subsubsection{Low-level Computational Blocks}
The baseline model is \textbf{VoxelMorph (VXM)}~(\cite{balakrishnan2019voxelmorph}), implemented as a vanilla U-Net~(\cite{ronneberger2015u}) with convolutional downsampling and transposed-convolution upsampling layers following~\cite{jian2024rethinkreg}. The encoder and decoder channel numbers are set to [16, 32, 64, 96, 128] and [128, 96, 64, 32], respectively, followed by two remaining convolutional layers (32 channels) before the \textit{Flow Convolution} layer that predicts the displacement field.

We then replace the convolutional encoder of the baseline with various computational backbones while keeping all other components identical:
- \textbf{Mam-VXM}: substitutes the CNN encoder with \textit{Mamba}~(\cite{gu2024mamba} blocks), while preserving the VXM decoder.
- \textbf{TransMorph (TM)}~(\cite{chen2022transmorph}): replaces the convolutional encoder with a Swin Transformer encoder. The original implementation\footnote{\url{https://github.com/junyuchen245/TransMorph_Transformer_for_Medical_Image_Registration}} (embedding dimension 96) is used.
- \textbf{Mam-TM}: identical to TransMorph except that Swin Transformer blocks are replaced with Mamba blocks.
- \textbf{LKU-Net}~(\cite{jia2022lku}): adds parallel large-kernel convolutions ($k=5$) to enlarge the receptive field. The original implementation\footnote{\url{https://github.com/xi-jia/LKU-Net}} (initial feature channel 16) is used.

\subsubsection{High-level Registration-specific Designs}
We build all variants on top of the canonical VXM backbone to isolate the effect of each registration-specific design component. The following modules are incrementally added:

\paragraph{Dual-stream Encoder (D)}
Instead of concatenating source and target images at the input, two parallel encoders with halved feature channels as VXM and shared weights, process them independently. The resulting feature pyramids are concatenated at each level before decoding. The pyramids consist of features at resolutions [1/16, 1/8, 1/4, 1/2].

\paragraph{Motion Pyramid and Warping (P)}
A hierarchical motion estimation scheme is introduced, producing deformation fields at multiple resolutions (1/16, 1/8, 1/4, 1/2). The remaining 2 convolution layers are removed, and the \textit{Flow Convolution} is added at each level $\ell$ to predict the multi-scale flow fields $\phi^\ell$. The low-resolution flow is upsampled and used to warp the moving feature map before the next refinement stage.

\paragraph{Correlation Volume (C)}
A localized correlation layer computes the inner product between source and target features within a local search window, forming a correlation volume. The target features serve as the context and are concatenated with the correlation volume, which together are passed to the decoder to enhance voxel-wise feature correspondence without explicit supervision. The search window of the correlation layer at each resolution level is set using a radius of $r=1$ for simplicity and efficiency.

\paragraph{Iterative Refinement (I)}
Iterative refinement applies recursive updates to the deformation field at a fixed resolution. The currently warped features are fed back into the decoder for an additional refinement pass. We set $n=2$ iterations at 1/4 and 1/2 resolutions following~\cite{jian2024rethinkreg}.

We intentionally use a consistent set of hyperparameters across all model variants rather than performing exhaustive, per-model tuning. The goal of this study is a \textit{ceteris paribus} (all other things being equal) comparison to isolate the impact of architectural choices, not to find the peak performance of each individual configuration. This ensures that any observed performance differences are attributable to the components being tested.

\subsection{Training Protocol}
All models are trained in a fully unsupervised manner using the training loss 
$\mathcal{L} = \mathcal{L}_\mathrm{sim} + \lambda \mathcal{L}_\mathrm{reg}$. 
We use the diffusion regularizer on the spatial gradients of the displacement $\mathcal{L}_\text{reg}=\|\nabla \mathbf{u}\|^2$ as the smoothness regularization, and select $\mathcal{L}_\mathrm{sim}$ and $\lambda$ based on the modality and task.

\paragraph{Pyramidal Training Loss}
For architectures containing motion pyramids, we compute losses at each resolution level $\ell$ and weight them proportionally to the scale:
\begin{equation}
\mathcal{L}_\mathrm{pyramid} = \sum_{\ell} 2^{-\ell} \left( 
\mathcal{L}_\mathrm{sim}(I_t^\ell, I_s^\ell \circ \phi^\ell) + 
\lambda \mathcal{L}_\mathrm{reg}(\phi^\ell)
\right),
\end{equation}
where downsampled images $I_t^\ell$ and $I_s^\ell$ are obtained by average pooling.

\paragraph{Optimization and Fairness}
All experiments use the Adam optimizer with an initial learning rate of $10^{-4}$ and an exponential decay rate of $0.996$. 
The random seed for both training and testing is fixed at 2023, ensuring that each method is trained on exactly the same randomly sampled image pairs per iteration. 
A bidirectional training strategy is adopted. Each training iteration performs both forward ($I_s \to I_t$) and backward ($I_t \to I_s$) registrations, with the total loss averaged across directions. 
This setting guarantees a fair, controlled comparison among all model variants.

\begin{table}[t]
\label{tab:model_profile}
    \caption{Architectural profiles of the evaluated methods. We report the number of trainable parameters (Params, in millions), floating-point operations (FLOPs, in GB), peak GPU memory usage during training (Max. Mem Train, in GB), and inference runtime per image pair (Runtime Infer, in seconds). All values are measured on images of size $160\times192\times224$.}
    \centering
    \setlength{\tabcolsep}{2.5pt} 
    \begin{tabular}{lcccc}
    \toprule
       &  Params & FLOPs & Max.Mem & Runtime\\
       &    (M)   & (GB) &  Train (GB) & Infer (s)\\
    \midrule
    VXM & 2.64 & 470 & 6.23 & 0.17\\
    Mam-VXM & 2.38 & 462 & 8.56 & 0.09\\
    TM & 46.56 & 612 & 10.24 & 0.20 \\
    Mam-TM & 46.77 & 590 & 8.02 & 0.06 \\
    LKU & 2.07 & 263 & 5.12 & 0.13 \\
    \midrule
    Dual & 1.92 & 433 & 5.13 & 0.28 \\
    DWP & 1.89 & 339 & 5.07 & 0.25 \\
    DWCP & 1.71 & 352 & 5.35 & 0.31 \\
    DWCPI & 1.71 & 646 & 5.35 & 0.50 \\
    \bottomrule
    \end{tabular}
\end{table}

\subsection{Evaluation Protocol}
We assess each model in terms of registration accuracy, deformation plausibility, and computational efficiency~(\cite{rohlfing2011unreliable}).

\paragraph{Registration Accuracy}
We compute:
(1) Dice similarity coefficient (DSC, \%), 
(2) 95th percentile Hausdorff distance (HD95), 
(3) average symmetric surface distance (ASSD), 
(4) normalized surface dice (NSD), and 
(5) target registration error (TRE) for keypoint annotations. 
Labels and keypoints are used exclusively for evaluation and not during training.

\paragraph{Deformation Smoothness}
We quantify the regularity of the deformation fields using:
(1) the standard deviation of the log-Jacobian determinant (SD$\log $J)~(\cite{leow2007sdlogj}), and 
(2) the ratio of non-diffeomorphic voxels (NDV)~(\cite{liu2024ndv}). 
Lower values indicate smoother and more physically plausible transformations.

\paragraph{Qualitative Visualization}
For qualitative comparison, we visualize: 
(1) target and warped source contours, 
(2) intensity difference maps for mono-modal registration, 
and (3) deformation grids. 
This allows visual assessment of alignment quality and the smoothness of local displacements.

\paragraph{Model Efficiency Profiling}
To decouple accuracy gains from increased computational cost, we report for each method:
(1) number of trainable parameters (Params, M), 
(2) number of floating-point operations (FLOPs, GB), 
(3) maximum GPU memory during training and inference (GB)

\subsection{Implementation Details}
All experiments are implemented in \texttt{PyTorch} 2.2 with CUDA 12.1 and trained using mixed precision. 
Training and inference are performed on NVIDIA A100 GPUs (40 GB memory). 
Each experiment is executed under the same environment, and all random seeds and preprocessing parameters are fixed to ensure full reproducibility.

We release all model configurations, training scripts, and evaluation code at \url{https://github.com/BailiangJ/rethink-reg} to facilitate transparent benchmarking.

\section{Results}\label{sec:results}
\subsection{Cross-sectional Brain MRI Registration}
\Cref{tab:lumir_lpba_mindboggle} and~\Cref{tab:oasis_adni_ixi} summarize the quantitative registration performance across six independent brain MRI datasets. Consistent with our previous findings in~\cite{jian2024rethinkreg}, even under an unsupervised training setup, trend-driven computational blocks, such as Mamba and large-kernel convolution, yield only marginal accuracy gains, or even slight degradation, over the VXM baseline. In contrast, high-level registration-specific designs yield substantially stronger and more consistent improvements across all datasets.

Among these designs, the dual-stream encoder (\textbf{D}) yields a modest Dice decrease, possibly due to its halved feature capacity. Introducing the motion pyramid and warping (\textbf{WP}) markedly boosts accuracy (e.g., +3-4\% DSC on LPBA) via coarse-to-fine refinement. Correlation volumes (\textbf{C}) further strengthen voxel-wise correspondence (+0.5-1\% DSC) and improve deformation plausibility, reflected by reduced SD$\log J$ and NDV.

Iterative refinement (\textbf{I}) provides a final nudge (e.g., +0.1-0.3\% DSC on OASIS and ADNI), indicating that recursive updates help capture subtle anatomical variations. However, this refinement occasionally increases SD$\log$J, as observed on OASIS, suggesting slightly less consistent local smoothness despite fewer foldings (NDV$\approx$0).

\begin{figure*}[t]
    \centering
    \includegraphics[width=1.0\linewidth]{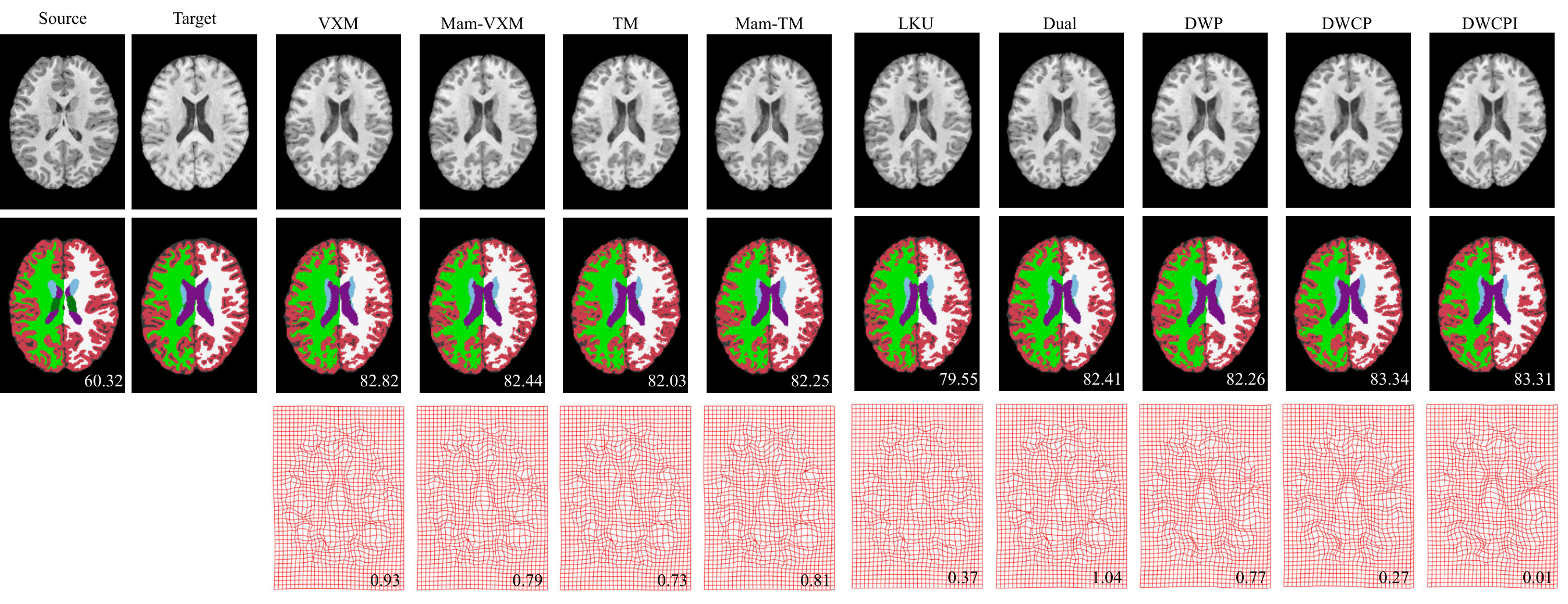}
    \caption{Qualitative registration results on the LUMIR25 in-distribution (ID) dataset. The axial views of the source, target, and registered images are shown alongside their corresponding segmentation labels. The deformation field is illustrated as deformed grid lines. The Dice similarity coefficient (DSC, \%) and the ratio of non-positive determinant voxels (NDV) across the volume are indicated in the bottom right corner.
    }
    \label{fig:lumir_qual}
    \vspace{0.5cm}
\end{figure*}

The qualitative visualizations in~\Cref{fig:lumir_qual} further corroborate these trends: incorporating registration-specific components cumulatively improves spatial alignment and yields more anatomically plausible deformations.

\begin{table*}[t]
    \scriptsize
    \caption{Quantitative results for cross-sectional brain registration on LUMIR 25 (In-Distribution (ID) (9 pairs) and Out-of-Distribution (OOD)) (18 pairs), LPBA (200 random pairs from 40 subjects), and MindBoggle (200 random pairs from 100 subjects). Accuracy is measured by Dice score (DSC, \%), 95th percentile Hausdorff distance (HD95), Target Registration Error (TRE), Deformation regularity is quantified by standard deviation of log-Jacobian determinant (SD$\log$J, $\times 10^{-2}$) and the ratio of non-diffeomorphic volume (NDV, \%). The best (\colorboxlegend{C1}{6pt}, \textbf{bold}), second best (\colorboxlegend{C2}{6pt}, \textit{italic}), and third best (\colorboxlegend{C3}{6pt}) results are highlighted.}
    \centering
    \setlength{\tabcolsep}{1.9pt} 
    \begin{tabular}{lcccccccccccccccc}
        \toprule
        & \multicolumn{6}{c}{LUMIR25} & \multicolumn{4}{c}{LPBA} & \multicolumn{4}{c}{MindBoggle} \\
        \cmidrule(lr){2-7} \cmidrule(lr){8-11} \cmidrule(lr){12-15}
        & \multicolumn{2}{c}{ID} & \multicolumn{4}{c}{OOD} & \multicolumn{4}{c}{} & \multicolumn{4}{c}{} \\
        \cmidrule(lr){2-3} \cmidrule(lr){4-7}
        & DSC $\uparrow$ & HD95 $\downarrow$ & DSC $\uparrow$ & HD95 $\downarrow$ & TRE $\downarrow$ & NDV(\%) $\downarrow$
        & DSC $\uparrow$ & HD95 $\downarrow$ & SD$\log$J $\downarrow$ & NDV(\%) $\downarrow$
        & DSC $\uparrow$ & HD95 $\downarrow$ & SD$\log$J $\downarrow$ & NDV(\%) $\downarrow$ \\
        \midrule
        Initial & 56.68 $\pm$ 1.67 & 4.79 $\pm$ 0.24 & 51.89 $\pm$ 3.62 & 5.27 $\pm$ 0.61 & 4.35 $\pm$ 0.55 & 0.00 $\pm$ 0.00 & 54.0$\pm$14.2 & 8.31$\pm$3.46 & - & - & 51.5$\pm$24.5 & 5.16$\pm$3.88 & - & - \\
        \midrule
        VXM & 74.35 $\pm$ 1.29 & 3.81 $\pm$ 0.21 & 70.26 $\pm$ 3.35 & 4.23 $\pm$ 0.62 & 2.64 $\pm$ 0.39 & 0.31 $\pm$ 0.10 & 67.1$\pm$13.6 & 7.60$\pm$3.48 & 7.2$\pm$0.5 & 1.5$\pm$0.7 & 67.2$\pm$24.1 & 5.47$\pm$4.75 & 7.6$\pm$0.4 & 0.8$\pm$0.2 \\
        Mam-VXM & 74.02 $\pm$ 1.75 & 3.77 $\pm$ 0.21 & 70.27 $\pm$ 3.22 & 4.21 $\pm$ 0.59 & 2.65 $\pm$ 0.37 & 0.25 $\pm$ 0.08 & 68.1$\pm$12.9 & 7.44$\pm$3.38 & 6.7$\pm$0.6 & 0.9$\pm$0.4 & 67.5$\pm$24.0 & 5.46$\pm$4.75 & \cellcolor{C3}7.4$\pm$0.3 & 0.8$\pm$0.1 \\
        TM & 74.06 $\pm$ 1.55 & 3.74 $\pm$ 0.20 & 69.59 $\pm$ 2.80 & 4.14 $\pm$ 0.52 & 2.71 $\pm$ 0.37 & \cellcolor{C3}0.23 $\pm$ 0.09 & 68.2$\pm$12.8 & 7.42$\pm$3.39 & 6.6$\pm$0.5 & 0.9$\pm$0.4 & 67.1$\pm$24.1 & 5.46$\pm$4.77 & 7.5$\pm$0.4 & 0.8$\pm$0.1 \\
        Mam-TM & 74.01 $\pm$ 1.71 & 3.73 $\pm$ 0.20 & 69.26 $\pm$ 3.01 & 4.22 $\pm$ 0.55 & 2.68 $\pm$ 0.40 & 0.26 $\pm$ 0.10 & 67.8$\pm$13.0 & 7.46$\pm$3.42 & 7.0$\pm$0.6 & 1.1$\pm$0.5 & 66.9$\pm$24.0 & 5.46$\pm$4.75 & 7.7$\pm$0.4 & 0.9$\pm$0.2 \\
        LKU & 74.35 $\pm$ 1.29 & 3.81 $\pm$ 0.21 & 70.26 $\pm$ 3.35 & 4.23 $\pm$ 0.62 & 2.64 $\pm$ 0.39 & 0.31 $\pm$ 0.10 & 65.8$\pm$13.2 & 7.65$\pm$3.37 & \cellcolor{C3}6.2$\pm$0.8 & 1.1$\pm$1.0 & 64.8$\pm$24.5 & 5.64$\pm$4.78 & 7.5$\pm$0.5 & 0.9$\pm$0.2 \\
        \midrule
        Dual & 74.37 $\pm$ 1.30 & 3.71 $\pm$ 0.20 & 70.14 $\pm$ 3.08 & 4.19 $\pm$ 0.57 & 2.73 $\pm$ 0.40 & 0.33 $\pm$ 0.08 & 67.3$\pm$13.0 & 7.51$\pm$3.38 & 7.1$\pm$0.7 & 1.2$\pm$0.5 & 67.1$\pm$24.0 & 5.47$\pm$4.74 & 7.8$\pm$0.5 & 0.9$\pm$0.1 \\
        DWP & \cellcolor{C3}76.44 $\pm$ 1.40 & \cellcolor{C1}\textbf{3.32 $\pm$ 0.19} & \cellcolor{C2}\textit{73.52 $\pm$ 1.90} & \cellcolor{C1}\textbf{3.54 $\pm$ 0.37} & \cellcolor{C3}2.51 $\pm$ 0.38 & 0.28 $\pm$ 0.07 & \cellcolor{C3}70.8$\pm$10.7 & \cellcolor{C3}6.89$\pm$3.05 & 6.3$\pm$0.6 & \cellcolor{C3}0.6$\pm$0.1 & \cellcolor{C2}\textit{67.7$\pm$24.1} & \cellcolor{C3}5.42$\pm$4.78 & 7.5$\pm$0.4 & \cellcolor{C3}0.8$\pm$0.1 \\
        DWCP & \cellcolor{C1}\textbf{76.90 $\pm$ 1.51} & \cellcolor{C2}\textit{3.32 $\pm$ 0.20} & \cellcolor{C3}73.39 $\pm$ 2.60 & \cellcolor{C2}\textit{3.65 $\pm$ 0.54} & \cellcolor{C2}\textit{2.42 $\pm$ 0.32} & \cellcolor{C2}\textit{0.09 $\pm$ 0.04} & \cellcolor{C1}\textbf{71.6$\pm$10.0} & \cellcolor{C1}\textbf{6.78$\pm$2.95} & \cellcolor{C1}\textbf{5.3$\pm$1.2} & \cellcolor{C2}\textit{0.1$\pm$0.1} & \cellcolor{C3}67.6$\pm$24.0 & \cellcolor{C1}\textbf{5.40$\pm$4.74} & \cellcolor{C1}\textbf{6.9$\pm$0.4} & \cellcolor{C2}\textit{0.2$\pm$0.1} \\
        DWCPI & \cellcolor{C2}\textit{76.88 $\pm$ 1.44} & \cellcolor{C3}3.34 $\pm$ 0.17 & \cellcolor{C1}\textbf{73.79 $\pm$ 2.57} & \cellcolor{C3}3.65 $\pm$ 0.58 & \cellcolor{C1}\textbf{2.36 $\pm$ 0.32} & \cellcolor{C1}\textbf{0.01 $\pm$ 0.01} & \cellcolor{C2}\textit{71.3$\pm$10.3} & \cellcolor{C2}\textit{6.85$\pm$2.99} & \cellcolor{C2}\textit{5.6$\pm$0.6} & \cellcolor{C1}\textbf{0.0$\pm$0.0} & \cellcolor{C1}\textbf{67.7$\pm$23.9} & \cellcolor{C2}\textit{5.40$\pm$4.77} & \cellcolor{C2}\textit{7.2$\pm$0.4} & \cellcolor{C1}\textbf{0.0$\pm$0.0} \\
        \bottomrule
    \end{tabular}
    \label{tab:lumir_lpba_mindboggle}
    \vspace{0.5cm}
\end{table*}

\begin{table*}[ht!]
\footnotesize
    \caption{Zero-shot cross-sectional brain MRI registration on OASIS (84 subjects), ADNI (43 subjects), and IXI (115 subjects) datasets, using 200 randomly sampled image pairs each. Metrics as in \Cref{tab:lumir_lpba_mindboggle}. The best (\colorboxlegend{C1}{6pt}, \textbf{bold}), second best (\colorboxlegend{C2}{6pt}, \textit{italic}), and third best (\colorboxlegend{C3}{6pt}) results are highlighted.}
    \centering
    \setlength{\tabcolsep}{3pt} 
    \begin{tabular}{lcccccccccccc}
        \toprule
         & \multicolumn{4}{c}{OASIS} & \multicolumn{4}{c}{ADNI} & \multicolumn{4}{c}{IXI} \\
        \cmidrule(lr){2-5} \cmidrule(lr){6-9} \cmidrule(lr){10-13}
         & DSC $\uparrow$ & HD95 $\downarrow$ & SD$\log$J $\downarrow$ & NDV(\%) $\downarrow$
         & DSC $\uparrow$ & HD95 $\downarrow$ & SD$\log$J $\downarrow$ & NDV(\%) $\downarrow$
         & DSC $\uparrow$ & HD95 $\downarrow$ & SD$\log$J $\downarrow$ & NDV(\%) $\downarrow$\\
        \midrule
        initial & 57.3$\pm$22.4 & 4.20$\pm$2.41 & - & - & 52.8$\pm$22.1 & 4.79$\pm$2.82 & - & - & 54.5$\pm$23.7 & 4.87$\pm$3.83 & - & - \\
        \midrule
        VXM & 79.7$\pm$17.7 & 5.14$\pm$5.63 & 7.9$\pm$0.5 & 0.9$\pm$0.2 & 74.0$\pm$18.7 & 4.64$\pm$3.93 & 8.0$\pm$0.4 & 1.1$\pm$0.3 & 71.7$\pm$21.3 & 5.21$\pm$5.38 & 8.0$\pm$0.4 & 1.1$\pm$0.3 \\
        Mam-VXM & 78.4$\pm$17.9 & 5.14$\pm$5.47 & 7.6$\pm$0.4 & 0.8$\pm$0.2 & 73.8$\pm$18.7 & 4.66$\pm$4.02 & 7.7$\pm$0.3 & 0.9$\pm$0.2 & 70.0$\pm$21.6 & 5.31$\pm$5.38 & 7.7$\pm$0.4 & 0.8$\pm$0.3 \\
        TM & 78.1$\pm$18.2 & 5.18$\pm$5.65 & \cellcolor{C2}\textit{7.3$\pm$0.4} & 0.7$\pm$0.2 & 73.2$\pm$18.8 & 4.69$\pm$4.05 & 7.6$\pm$0.4 & 0.9$\pm$0.3 & 70.2$\pm$21.6 & 5.31$\pm$5.48 & \cellcolor{C3}7.4$\pm$0.4 & 0.8$\pm$0.3 \\
        Mam-TM & 78.2$\pm$18.1 & 5.15$\pm$5.48 & \cellcolor{C2}\textit{7.3$\pm$0.4} & 0.8$\pm$0.2 & 72.9$\pm$18.9 & 4.70$\pm$4.00 & 7.8$\pm$0.5 & 1.1$\pm$0.3 & 69.9$\pm$21.8 & 5.30$\pm$5.41 & \cellcolor{C2}\textit{7.3$\pm$0.4} & 0.8$\pm$0.4 \\
        LKU & 74.1$\pm$20.1 & 5.29$\pm$5.49 & \cellcolor{C1}\textbf{6.3$\pm$0.4} & \cellcolor{C2}\textit{0.2$\pm$0.1} & 70.0$\pm$20.0 & 4.84$\pm$3.94 & \cellcolor{C1}\textbf{6.7$\pm$0.4} & \cellcolor{C3}0.5$\pm$0.3 & 67.4$\pm$22.6 & 5.43$\pm$5.36 & \cellcolor{C1}\textbf{6.3$\pm$0.5} & \cellcolor{C2}\textit{0.4$\pm$0.3} \\
        \midrule
        Dual & 78.6$\pm$18.1 & 5.19$\pm$5.62 & 8.2$\pm$0.5 & 1.1$\pm$0.2 & 73.5$\pm$19.0 & 4.66$\pm$3.95 & 8.3$\pm$0.4 & 1.2$\pm$0.3 & 70.6$\pm$21.7 & 5.29$\pm$5.45 & 8.2$\pm$0.4 & 1.1$\pm$0.3 \\
        DWP & \cellcolor{C3}80.0$\pm$17.6 & \cellcolor{C3}5.08$\pm$5.53 & 7.7$\pm$0.5 & 0.9$\pm$0.2 & \cellcolor{C3}75.2$\pm$18.3 & \cellcolor{C3}4.58$\pm$4.03 & 7.7$\pm$0.4 & 0.9$\pm$0.2 & \cellcolor{C3}72.4$\pm$21.1 & \cellcolor{C3}5.19$\pm$5.51 & 8.0$\pm$0.6 & 1.0$\pm$0.2 \\
        DWCP & \cellcolor{C2}\textit{80.9$\pm$17.0} & \cellcolor{C2}\textit{5.03$\pm$5.57} & 9.6$\pm$2.6 & \cellcolor{C3}0.3$\pm$0.1 & \cellcolor{C2}\textit{75.5$\pm$18.1} & \cellcolor{C2}\textit{4.55$\pm$4.03} & \cellcolor{C2}\textit{7.2$\pm$0.4} & \cellcolor{C2}\textit{0.4$\pm$0.1} & \cellcolor{C2}\textit{73.1$\pm$20.9} & \cellcolor{C1}\textbf{5.10$\pm$5.40} & 7.6$\pm$0.8 & \cellcolor{C3}0.5$\pm$0.2 \\
        DWCPI & \cellcolor{C1}\textbf{81.1$\pm$16.8} & \cellcolor{C1}\textbf{5.03$\pm$5.56} & 10.2$\pm$3.1 & \cellcolor{C1}\textbf{0.1$\pm$0.0} & \cellcolor{C1}\textbf{75.6$\pm$18.0} & \cellcolor{C1}\textbf{4.54$\pm$4.01} & \cellcolor{C3}7.4$\pm$0.4 & \cellcolor{C1}\textbf{0.0$\pm$0.0} & \cellcolor{C1}\textbf{73.2$\pm$20.7} & \cellcolor{C2}\textit{5.10$\pm$5.42} & 7.7$\pm$0.5 & \cellcolor{C1}\textbf{0.0$\pm$0.0} \\
        \bottomrule
    \end{tabular}
    \label{tab:oasis_adni_ixi}
    \vspace{0.5cm}
\end{table*}

\subsection{Longitudinal Lung CT Registration}
\begin{figure*}[ht]
    \centering
    \includegraphics[width=0.95\linewidth]{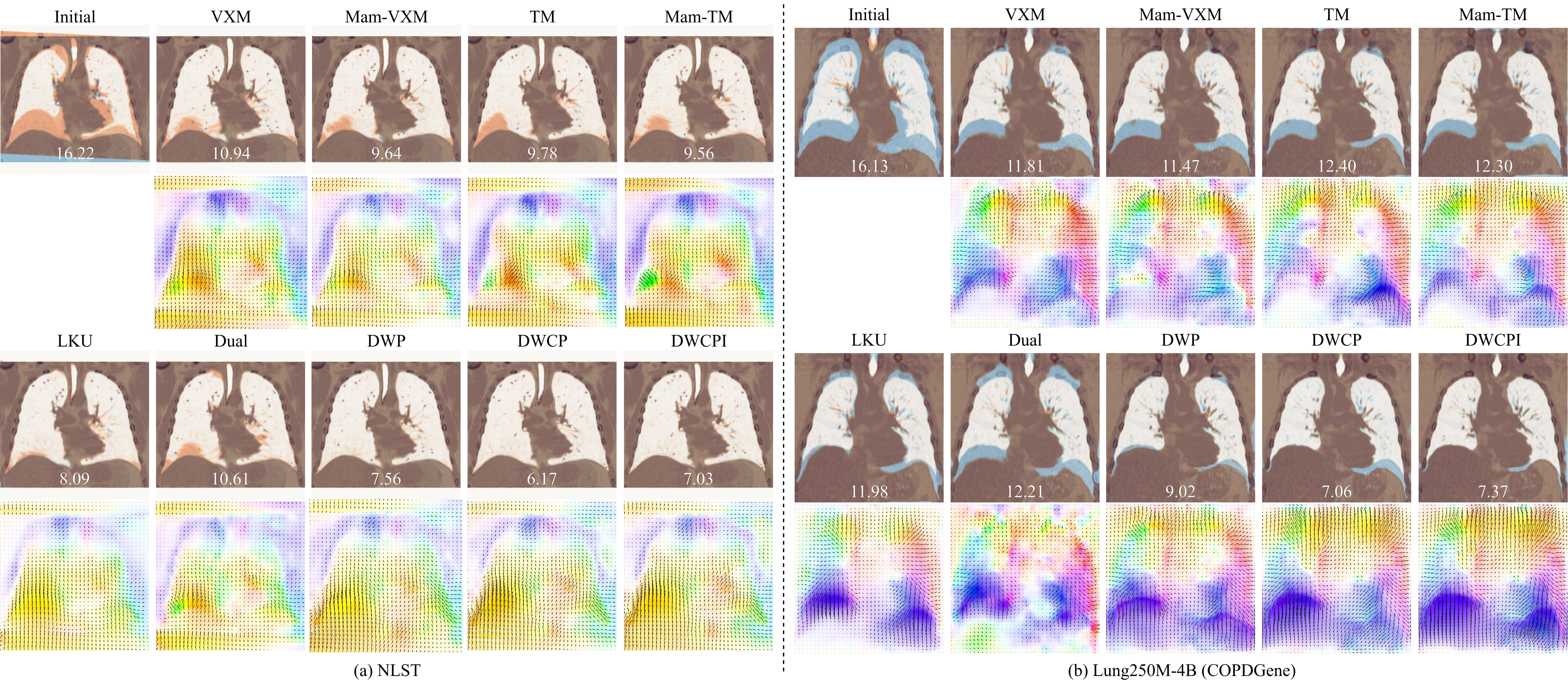}
    \caption{
    Qualitative registration results on the NLST (in-domain) and Lung250M-4B (zero-shot) datasets. The coronal overlays of the registered moving and target images are shown alongside their corresponding deformation fields. The target and source images are visualized in \textbf{blue} and \textbf{organ}, and a perfectly aligned lung appears \textbf{white} in the overlay. The deformation field is visualized by a colored quiver plot where the color and arrow direction encode the displacement orientation, and arrow length represents displacement magnitude. The target registration error (TRE) for each case is indicated at the bottom of the image.
    }
    \label{fig:lung_qual_hor}
    \vspace{-0.15cm}
\end{figure*}

\begin{table*}[ht]
    \scriptsize
    \centering
    \caption{Quantitative results on the NLST (in-domain) and Lung250M-4B (zero-shot) datasets. Reported metrics include target registration error (TRE, mm or vox), Dice score (DSC, \%), and deformation regularity measures: SD$\log$J, ($\times 10^{-2}$); and the percentage of non-diffeomorphic voxels (NDV) across the image volume. The best (\colorboxlegend{C1}{6pt}, \textbf{bold}), second best (\colorboxlegend{C2}{6pt}, \textit{italic}), and third best (\colorboxlegend{C3}{6pt}) results are highlighted.}
    \label{tab:lung}
    \setlength{\tabcolsep}{2.75pt}
    \begin{tabular}{lcccccccccccccc}
        \toprule
        & \multicolumn{8}{c}{NLST (In-domain)} & \multicolumn{6}{c}{Lung250M-4B (Zero-shot)}\\
        \cmidrule(lr){2-9} \cmidrule(lr){10-15}
        & \multicolumn{4}{c}{INP $\rightarrow$ EXP} & \multicolumn{4}{c}{EXP $\rightarrow$ INP} & \multicolumn{3}{c}{INP $\rightarrow$ EXP} & \multicolumn{3}{c}{EXP $\rightarrow$ INP}\\
        \cmidrule(lr){2-5} \cmidrule(lr){6-9} \cmidrule(lr){10-12} \cmidrule(lr){13-15}
        & TRE $\downarrow$ & DSC $\uparrow$ & SD$\log$J $\downarrow$ & NDV(\%) $\downarrow$ & TRE $\downarrow$ & DSC $\uparrow$ & SD$\log$J $\downarrow$ & NDV(\%) $\downarrow$ & TRE $\downarrow$ & SD$\log$J $\downarrow$ & NDV(\%) $\downarrow$ & TRE $\downarrow$ & SD$\log$J $\downarrow$ & NDV(\%) $\downarrow$\\
        \midrule
        Initial & 10.14$\pm$2.92 & 89.2$\pm$3.0 & - & - & 10.14$\pm$2.92 & 89.2$\pm$3.0 & - & - & 13.74$\pm$3.63 & - & - & 13.74$\pm$3.63 & - & -\\
        \midrule
        VXM & 3.66$\pm$2.60 & 97.5$\pm$2.1 & \cellcolor{C2}\textit{7.4$\pm$1.3} & 0.58$\pm$0.78 & 3.62$\pm$2.59 & 97.6$\pm$1.9 & \cellcolor{C2}\textit{7.1$\pm$1.0} & 0.48$\pm$0.55 & 8.98$\pm$4.52 & \cellcolor{C2}\textit{10.5$\pm$2.8} & 0.72$\pm$0.65 & 8.86$\pm$4.38 & \cellcolor{C2}\textit{9.1$\pm$1.9} & 0.66$\pm$0.69\\
        Mam-VXM & 3.47$\pm$2.60 & 97.6$\pm$2.1 & 7.7$\pm$1.4 & 0.66$\pm$0.91 & 3.42$\pm$2.47 & 97.7$\pm$1.9 & 7.3$\pm$1.0 & 0.43$\pm$0.40 & 9.18$\pm$4.77 & 11.3$\pm$3.4 & 1.05$\pm$0.99 & 8.89$\pm$4.53 & 9.5$\pm$2.1 & 0.68$\pm$0.96\\
        TM & 3.41$\pm$2.53 & 97.6$\pm$2.0 & 7.6$\pm$1.2 & 0.55$\pm$0.63 & 3.46$\pm$2.61 & 97.6$\pm$2.1 & 7.4$\pm$1.1 & 0.65$\pm$0.80 & 9.38$\pm$4.94 & 11.4$\pm$3.5 & 1.11$\pm$0.97 & 8.87$\pm$4.60 & 9.3$\pm$2.1 & 0.66$\pm$0.60\\
        Mam-TM & 3.42$\pm$2.46 & 97.6$\pm$1.9 & \cellcolor{C2}\textit{7.4$\pm$1.3} & 0.48$\pm$0.56 & 3.38$\pm$2.46 & 97.7$\pm$1.8 & \cellcolor{C2}\textit{7.1$\pm$1.0} & 0.38$\pm$0.38 & 9.06$\pm$4.72 & \cellcolor{C3}11.0$\pm$3.6 & 1.08$\pm$1.01 & 8.88$\pm$4.51 & \cellcolor{C3}9.2$\pm$2.2 & 0.82$\pm$0.86\\
        LKU & 3.78$\pm$1.87 & 97.6$\pm$1.3 & \cellcolor{C1}\textbf{6.7$\pm$1.4} & 0.37$\pm$0.58 & 3.75$\pm$1.78 & 97.6$\pm$1.1 & \cellcolor{C1}\textbf{6.6$\pm$1.2} & \cellcolor{C3}0.19$\pm$0.24 & \cellcolor{C3}8.69$\pm$4.27 & \cellcolor{C1}\textbf{8.6$\pm$2.6} & \cellcolor{C3}0.21$\pm$0.27 & \cellcolor{C2}\textit{7.64$\pm$3.75} & \cellcolor{C1}\textbf{7.7$\pm$1.8} & \cellcolor{C3}0.18$\pm$0.22\\
        \midrule
        Dual & 4.41$\pm$2.37 & 97.4$\pm$2.1 & 7.8$\pm$1.3 & 0.50$\pm$0.53 & 4.42$\pm$2.48 & 97.5$\pm$2.1 & 7.5$\pm$1.0 & 0.44$\pm$0.48 & 11.27$\pm$4.12 & 11.6$\pm$1.9 & 0.99$\pm$0.75 & 11.23$\pm$4.13 & 10.4$\pm$1.6 & 0.83$\pm$0.70\\
        DWP & \cellcolor{C3}2.60$\pm$1.44 & \cellcolor{C1}\textbf{98.4$\pm$0.7} & 7.9$\pm$1.5 & \cellcolor{C3}0.31$\pm$0.38 & \cellcolor{C3}2.60$\pm$1.41 & \cellcolor{C1}\textbf{98.4$\pm$0.6} & 7.9$\pm$1.5 & 0.20$\pm$0.16 & 9.64$\pm$4.22 & 13.1$\pm$2.7 & 0.52$\pm$0.35 & 9.51$\pm$4.16 & 10.9$\pm$2.1 & 0.52$\pm$0.39\\
        DWCP & \cellcolor{C2}\textit{2.00$\pm$1.45} & \cellcolor{C3}98.4$\pm$1.0 & \cellcolor{C3}7.5$\pm$1.2 & \cellcolor{C2}\textit{0.13$\pm$0.12} & \cellcolor{C2}\textit{2.03$\pm$1.46} & \cellcolor{C2}\textit{98.4$\pm$0.7} & \cellcolor{C3}7.6$\pm$1.5 & \cellcolor{C2}\textit{0.11$\pm$0.10} & \cellcolor{C1}\textbf{8.04$\pm$4.02} & 12.7$\pm$3.1 & \cellcolor{C2}\textit{0.17$\pm$0.12} & \cellcolor{C1}\textbf{7.58$\pm$3.91} & 9.5$\pm$1.7 & \cellcolor{C2}\textit{0.13$\pm$0.11}\\
        DWCPI & \cellcolor{C1}\textbf{1.95$\pm$1.47} & \cellcolor{C2}\textit{98.4$\pm$0.9} & 7.6$\pm$1.1 & \cellcolor{C1}\textbf{0.05$\pm$0.06} & \cellcolor{C1}\textbf{2.00$\pm$1.49} & \cellcolor{C1}\textbf{98.4$\pm$0.6} & 7.7$\pm$1.3 & \cellcolor{C1}\textbf{0.04$\pm$0.05} & \cellcolor{C2}\textit{8.42$\pm$4.37} & 12.6$\pm$2.7 & \cellcolor{C1}\textbf{0.06$\pm$0.07} & \cellcolor{C3}7.88$\pm$4.31 & 9.6$\pm$1.4 & \cellcolor{C1}\textbf{0.02$\pm$0.02}\\
        \bottomrule
    \end{tabular}
    \vspace{-0.15cm}
\end{table*}
\begin{figure*}[h!]
    \centering
    \includegraphics[width=0.91\linewidth]{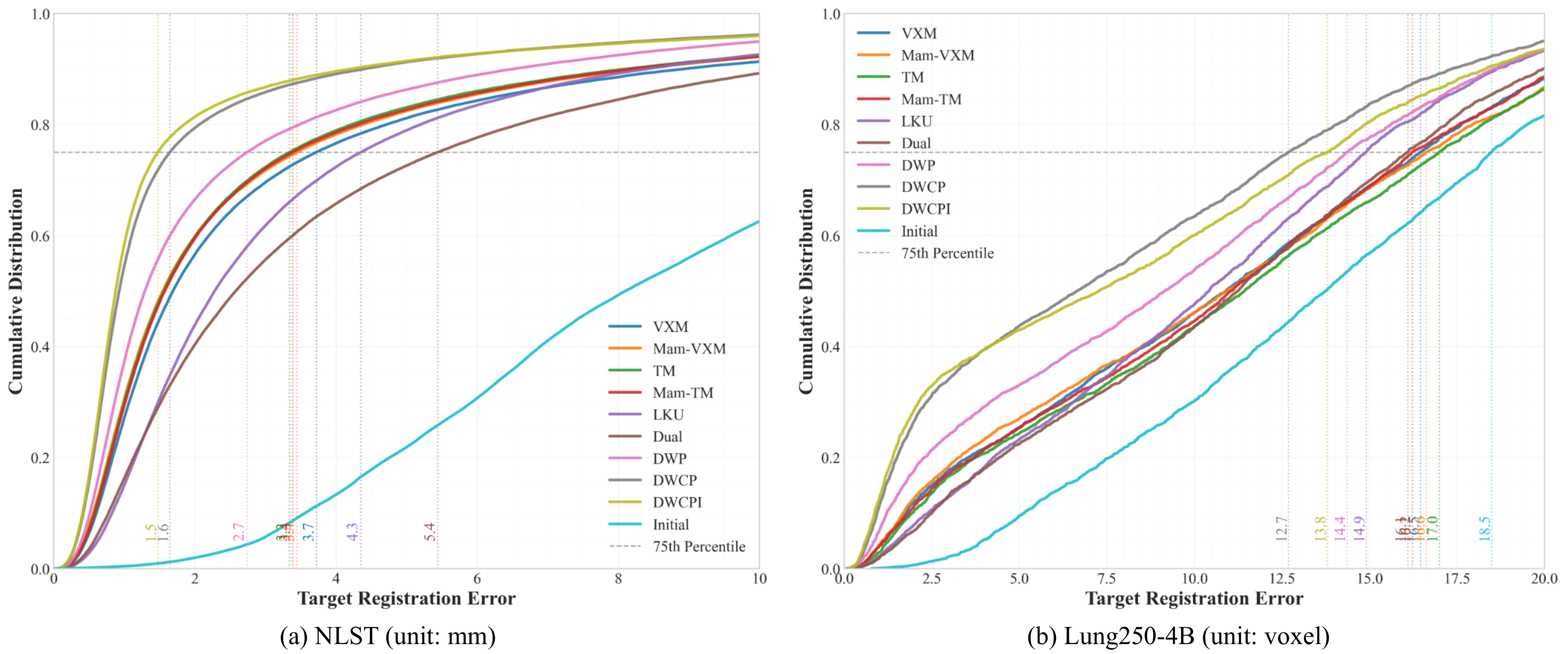}
    \caption{Cumulative distribution of target registration error (TRE) on the NLST (in-domain) and Lung250M-4B (zero-shot) datasets. The dotted vertical lines indicate the 75th percentile of TRE in each distribution. A larger area under the curve corresponds to better keypoints/landmarks alignment performance.
    }
    \label{fig:cum_tre}
\end{figure*}

\begin{figure*}[ht!]
    \centering
    \includegraphics[width=0.9\linewidth]{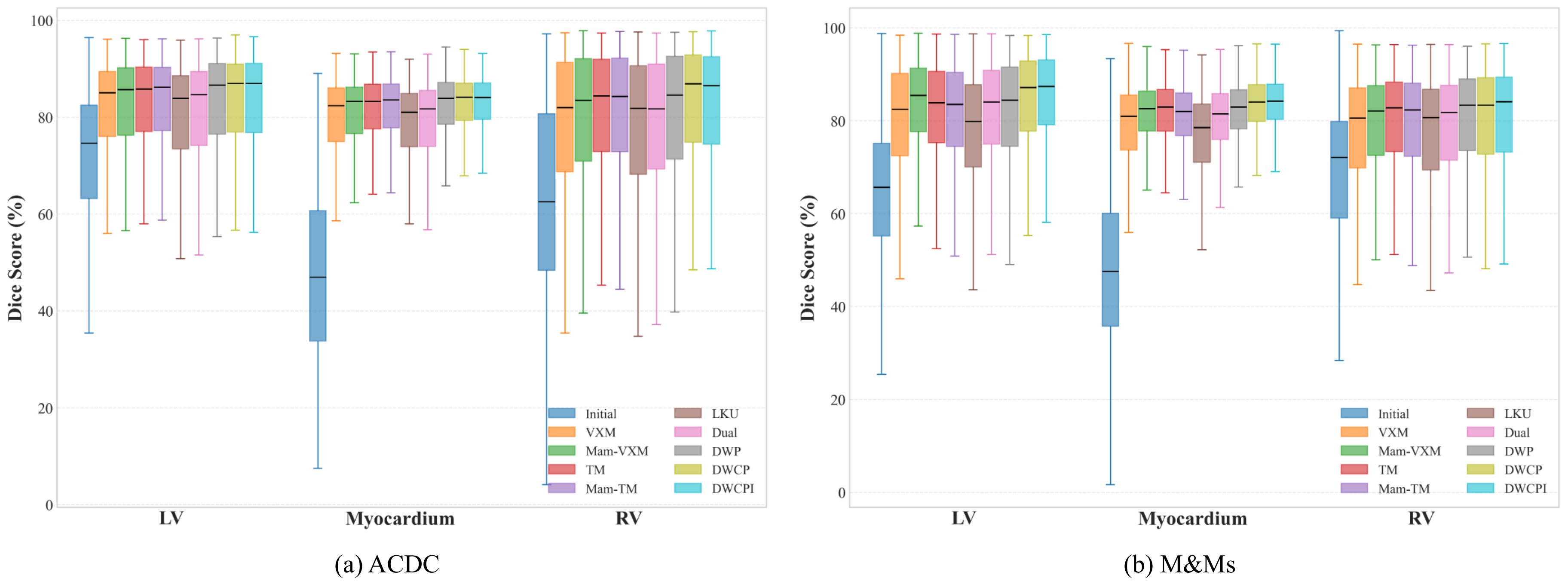}
    \caption{Box plots of structure-wise Dice scores on the ACDC (in-domain) and M\&Ms (zero-shot) datasets. Structures include the left ventricle (LV), myocardium (Myo), and right ventricle (RV).}
    \label{fig:cardiac_box}
    \vspace{-4pt}
\end{figure*}

\Cref{tab:lung} summarizes the results for intra-subject lung registration on the NLST test set and the zero-shot Lung250M-4B dataset. Overall, the results exhibit consistent trends across both in-domain and out-of-distribution (OOD) evaluations. Compared to the baseline CNN-based VoxelMorph (VXM), Transformer- and Mamba-based models achieve marginally improved alignment accuracy on the NLST test set, suggesting that their ability to capture long-range dependencies can better accommodate large respiratory motions. However, these gains are modest and come at a substantially higher computational cost. As illustrated in~\Cref{fig:cum_tre}, registration-specific designs (\textbf{DWP}, \textbf{DWCP}, \textbf{DWCPI}) consistently achieve markedly lower 75th percentile in the cumulative TRE distribution, indicating more accurate alignment for the majority of anatomical keypoints.

In contrast, registration-specific architectures yield more pronounced and systematic improvements. Incorporating the motion pyramid and warping modules (\textbf{WP}) reduces the mean TRE from 3.6 mm to 2.6 mm on NLST, accompanied by a +1\% increase in DSC, demonstrating that the coarse-to-fine refinement efficiently captures large deformations. Adding a correlation volume (\textbf{C}) further strengthens voxel-level correspondence, lowering the TRE to approximately 2.0 mm while improving deformation regularity. Notably, this design also generalizes best to the unseen Lung250M-4B dataset, where the \textbf{DWCP} variant achieves the lowest TRE and most plausible deformations, confirming the robustness of local correlation-based matching under domain shift.

The iterative refinement module (\textbf{I}) provides marginal gains on the in-domain NLST set but yields slightly degraded keypoint accuracy in the OOD scenario. Interestingly, although \textbf{DWCPI} exhibits slightly higher TRE than \textbf{DWCP} on Lung250M-4B, qualitative inspection in~\Cref{fig:lung_qual_hor} reveals the best alignment in the left lung. This discrepancy might suggest that the iterative refinement stage encourages deformations that better preserve overall shape and boundary continuity, while relaxing on the localized keypoint correspondences. A similar effect is observed in LKU-Net, which gets weaker TRE results but better overall lung alignment. This is likely due to its enlarged receptive field that promotes smooth, large-scale alignment at the cost of reduced local precision.

In summary, while long-range models (Transformers, Mamba) modestly improve within-domain performance, the most reliable accuracy–regularity trade-off and generalization are consistently achieved by the registration-specific priors, particularly the pyramidal and correlation-based designs.

\subsection{Temporal Cardiac MRI 2D Registration}
\begin{figure*}[ht]
    \centering
    \includegraphics[width=1.0\linewidth]{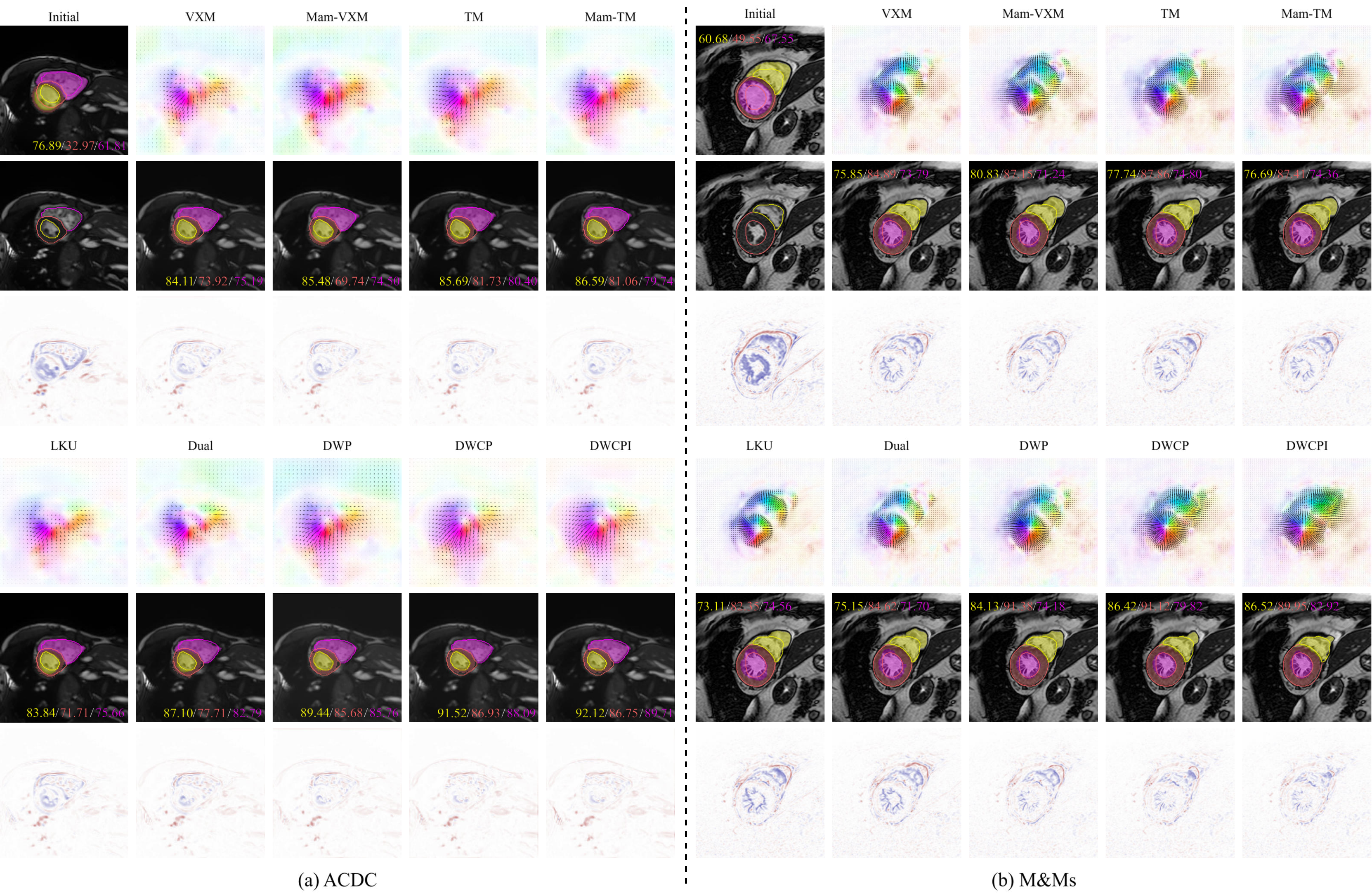}
    \caption{Qualitative registration results on the ACDC and M\&Ms dataset. The target image is shown with its segmentation contour overlaid. The source and registered images are overlaid with both their respective segmentation labels and the target contour. Subtraction error maps between the target and source/registered images are displayed, where white denotes zero intensity difference. Perfect registration aligns the contour precisely with the segmentation boundary. Dice scores for the left ventricle, myocardium, and right ventricle are reported within each image. The deformation field is visualized by a colored quiver plot, where the color and arrow direction encode the displacement orientation, and arrow length represents displacement magnitude.
    }\label{fig:cardiac_qual}
    \vspace{-4pt}
\end{figure*}
\begin{table*}[ht!]
    \centering
    \scriptsize
    \caption{Quantitative results on the ACDC (in-domain) and M\&Ms (zero-shot) test sets. Reported metrics include Dice score (DSC,\%), Normalized Surface Dice (NSD,\%), the standard deviation of the log-Jacobian determinant (SD$\log$J,$\times10^{-2}$), and the number of non-diffeomorphic voxels within the foreground (NDV,\textpertenthousand, per 10,000 voxels). The best (\colorboxlegend{C1}{6pt}, \textbf{bold}), second best (\colorboxlegend{C2}{6pt}, \textit{italic}), and third best (\colorboxlegend{C3}{6pt}) results are highlighted.}
    \label{tab:ACDC_mms}
    \setlength{\tabcolsep}{0.9pt} 
    \begin{tabular}{lcccccccccccccccc}
    \toprule
    & \multicolumn{8}{c}{ACDC} & \multicolumn{8}{c}{M\&Ms (Zero-shot)}\\
    \cmidrule(lr){2-9} \cmidrule(lr){10-17}
    & \multicolumn{4}{c}{ED $\rightarrow$ ES} & \multicolumn{4}{c}{ES $\rightarrow$ ED} & \multicolumn{4}{c}{ED $\rightarrow$ ES} & \multicolumn{4}{c}{ES $\rightarrow$ ED}\\
    \cmidrule(lr){2-5} \cmidrule(lr){6-9} \cmidrule(lr){10-13} \cmidrule(lr){14-17}
    & DSC $\uparrow$ & NSD $\uparrow$ & SD$\log$J $\downarrow$ & NDV(\textpertenthousand) $\downarrow$ & DSC $\uparrow$ & NSD $\uparrow$ & SD$\log$J $\downarrow$ & NDV(\textpertenthousand) $\downarrow$ & DSC $\uparrow$ & NSD $\uparrow$ & SD$\log$J $\downarrow$ & NDV(\textpertenthousand) $\downarrow$ & DSC $\uparrow$ & NSD $\uparrow$ & SD$\log$J $\downarrow$ & NDV(\textpertenthousand) $\downarrow$\\
    \midrule
    initial & 60.2$\pm$14.8 & 10.7$\pm$7.3 & - & - & 60.2$\pm$14.8 & 10.7$\pm$7.3 & - & - & 60.4$\pm$12.7 & 7.9$\pm$6.0 & - & - & 60.4$\pm$12.7 & 7.9$\pm$6.0 & - & -\\
    \midrule
    VXM & 78.8$\pm$10.8 & 22.4$\pm$6.2 & \cellcolor{C3}4.7$\pm$1.6 & 0.3$\pm$1.8 & 83.9$\pm$8.3 & 20.8$\pm$7.2 & \cellcolor{C3}4.1$\pm$1.2 & 3.0$\pm$13.3 & 77.7$\pm$11.0 & 15.1$\pm$6.3 & 4.2$\pm$1.8 & 16.8$\pm$40.5 & 83.8$\pm$7.8 & 18.8$\pm$7.7 & 3.4$\pm$1.0 & 15.5$\pm$34.9\\
    Mam-VXM & 80.0$\pm$9.9 & 22.8$\pm$6.1 & 5.1$\pm$1.9 & 1.5$\pm$5.6 & 85.3$\pm$7.6 & 22.5$\pm$7.4 & 4.4$\pm$1.5 & 7.0$\pm$21.0 & \cellcolor{C3}80.4$\pm$8.5 & 16.3$\pm$6.2 & 4.1$\pm$1.7 & 29.9$\pm$60.8 & \cellcolor{C3}85.0$\pm$7.1 & \cellcolor{C2}\textit{20.2$\pm$7.6} & 3.5$\pm$1.4 & 48.9$\pm$88.4\\
    TM & \cellcolor{C3}80.4$\pm$9.7 & 23.2$\pm$5.9 & 5.0$\pm$1.9 & 1.1$\pm$5.1 & \cellcolor{C3}85.5$\pm$7.4 & 22.3$\pm$7.2 & 4.3$\pm$1.4 & 4.4$\pm$17.9 & 80.2$\pm$9.1 & 17.9$\pm$6.4 & 3.4$\pm$1.5 & 1.9$\pm$6.8 & 84.8$\pm$7.5 & 17.9$\pm$8.2 & \cellcolor{C3}2.9$\pm$1.2 & 3.2$\pm$9.1\\
    Mam-TM & 80.4$\pm$9.9 & 22.9$\pm$6.2 & 5.1$\pm$1.9 & 1.2$\pm$5.4 & 85.4$\pm$7.7 & 22.3$\pm$7.1 & 4.3$\pm$1.4 & 5.4$\pm$18.6 & 79.5$\pm$9.3 & 18.0$\pm$6.6 & \cellcolor{C3}3.4$\pm$1.4 & 9.2$\pm$26.8 & 84.6$\pm$7.4 & 17.5$\pm$7.9 & \cellcolor{C2}\textit{2.9$\pm$1.1} & 10.3$\pm$25.0\\
    LKU & 77.7$\pm$10.7 & 21.6$\pm$6.4 & \cellcolor{C1}\textbf{4.5$\pm$1.9} & \cellcolor{C2}\textit{0.1$\pm$0.7} & 82.2$\pm$8.7 & 18.0$\pm$6.7 & \cellcolor{C1}\textbf{3.8$\pm$1.4} & 5.3$\pm$19.7 & 76.4$\pm$10.3 & 16.3$\pm$6.2 & \cellcolor{C1}\textbf{3.0$\pm$1.4} & \cellcolor{C1}\textbf{0.3$\pm$2.0} & 80.5$\pm$9.5 & 14.5$\pm$7.3 & \cellcolor{C1}\textbf{2.5$\pm$1.1} & 2.4$\pm$7.2\\
    \midrule
    Dual & 78.1$\pm$11.0 & 22.3$\pm$5.6 & \cellcolor{C2}\textit{4.7$\pm$1.5} & \cellcolor{C3}0.2$\pm$1.7 & 83.1$\pm$8.8 & 19.7$\pm$6.7 & \cellcolor{C2}\textit{3.9$\pm$1.1} & 2.7$\pm$12.8 & 79.3$\pm$9.5 & 17.4$\pm$6.1 & \cellcolor{C2}\textit{3.4$\pm$1.1} & 3.1$\pm$14.2 & 83.6$\pm$7.7 & 17.4$\pm$7.7 & \cellcolor{C2}\textit{2.9$\pm$1.1} & 3.5$\pm$11.9\\
    DWP & 80.3$\pm$10.9 & \cellcolor{C3}23.3$\pm$5.7 & 4.9$\pm$1.6 & \cellcolor{C1}\textbf{0.1$\pm$0.6} & 85.4$\pm$8.5 & \cellcolor{C3}23.2$\pm$7.5 & 4.3$\pm$1.3 & \cellcolor{C2}\textit{0.7$\pm$3.9} & 80.3$\pm$9.5 & \cellcolor{C3}18.8$\pm$6.3 & 3.5$\pm$1.2 & \cellcolor{C2}\textit{0.7$\pm$3.6} & \cellcolor{C2}\textit{85.3$\pm$7.6} & 18.8$\pm$7.8 & 3.2$\pm$1.2 & \cellcolor{C1}\textbf{0.8$\pm$3.3}\\
    DWCP & \cellcolor{C1}\textbf{81.3$\pm$10.3} & \cellcolor{C1}\textbf{23.9$\pm$5.7} & 5.2$\pm$3.5 & 0.5$\pm$4.1 & \cellcolor{C1}\textbf{86.1$\pm$7.9} & \cellcolor{C1}\textbf{24.4$\pm$7.8} & 4.4$\pm$1.3 & \cellcolor{C3}1.0$\pm$5.1 & \cellcolor{C2}\textit{81.3$\pm$9.6} & \cellcolor{C1}\textbf{19.0$\pm$6.3} & 3.9$\pm$3.5 & 1.6$\pm$9.5 & \cellcolor{C1}\textbf{86.2$\pm$7.5} & \cellcolor{C1}\textbf{20.3$\pm$7.5} & 3.5$\pm$3.7 & \cellcolor{C3}1.6$\pm$8.6\\
    DWCPI & \cellcolor{C2}\textit{81.2$\pm$10.4} & \cellcolor{C2}\textit{23.8$\pm$5.7} & 5.0$\pm$1.7 & 0.3$\pm$2.5 & \cellcolor{C2}\textit{85.8$\pm$8.4} & \cellcolor{C2}\textit{23.6$\pm$7.9} & 4.4$\pm$1.3 & \cellcolor{C1}\textbf{0.2$\pm$1.9} & \cellcolor{C1}\textbf{81.7$\pm$9.5} & \cellcolor{C1}\textbf{19.0$\pm$6.3} & 4.3$\pm$5.3 & \cellcolor{C3}1.3$\pm$7.8 & \cellcolor{C1}\textbf{86.2$\pm$7.4} & \cellcolor{C3}20.1$\pm$7.5 & 3.9$\pm$5.3 & \cellcolor{C2}\textit{1.2$\pm$7.2}\\
    \bottomrule
    \end{tabular}
    \vspace{-4pt}
\end{table*}

\Cref{tab:ACDC_mms} summarize results for the 2D cardiac MRI registration. The ACDC dataset represents the in-domain evaluation, while M\&Ms serves as a challenging zero-shot generalization test involving data from different scanners, countries, and patient pathologies. The task demands accurate alignment between end-diastolic (ED) and end-systolic (ES) phases, which involves complex non-rigid motions of the left ventricle (LV), right ventricle (RV), and myocardium.

Compared to convolutional baselines, Transformer- and Mamba-based models achieve notable gains in both Dice and NSD metrics (+1.5--3\% DSC and +0.5--1\% NSD). These improvements reflect their capability to capture long-range dependencies that better model the global contraction-expansion dynamics of cardiac motion. However, LKU-Net with large kernel convolutions exhibits slightly lower accuracy despite smoother deformation fields, suggesting that the enlarged receptive field may come at the cost of reduced sensitivity to local motion details critical for anatomical alignment.

Nonetheless, these trend-driven improvements are ultimately matched and surpassed by the integration of registration-specific designs. The motion pyramid combined with warping (\textbf{DWP}) substantially enhances performance through coarse-to-fine refinement (+$\sim 2$\% DSC), reaching the same level of performance as the Transformer baseline (TM) with greater efficiency. Adding a correlation volume (\textbf{DWCP}) further improves pixel-wise correspondence, yielding the best balance between alignment accuracy and deformation regularity (highest DSC with near-zero NDV). The explicit modeling of local similarity via the correlation enables more precise delineation of ventricular boundaries and myocardial contours (\Cref{fig:cardiac_qual}), particularly in regions with limited texture contrast. This advantage underscores that local correlation remains a powerful inductive bias, surpassing computation blocks capable of global context modeling.

While the iterative refinement variant (\textbf{DWCPI}) gets modest gains over \textbf{DWCP} on accuracy, qualitative inspection in~\Cref{fig:cardiac_qual} reveals that \textbf{DWCPI} achieves more accurate alignment of RV in the difficult M\&Ms case with significant shape changes. This highlights the potential of iterative refinement to enhance robustness in extreme motion scenarios.

Last but not least, as illustrated by the box plots in~\Cref{fig:cardiac_box}, registration-specific designs consistently yield higher and more stable DSC across all cardiac structures (LV, Myo, RV), confirming that the explicit registration priors play a decisive role in achieving cross-domain stability and anatomical consistency.

\begin{figure*}[ht!]
    \centering
    \includegraphics[width=\linewidth]{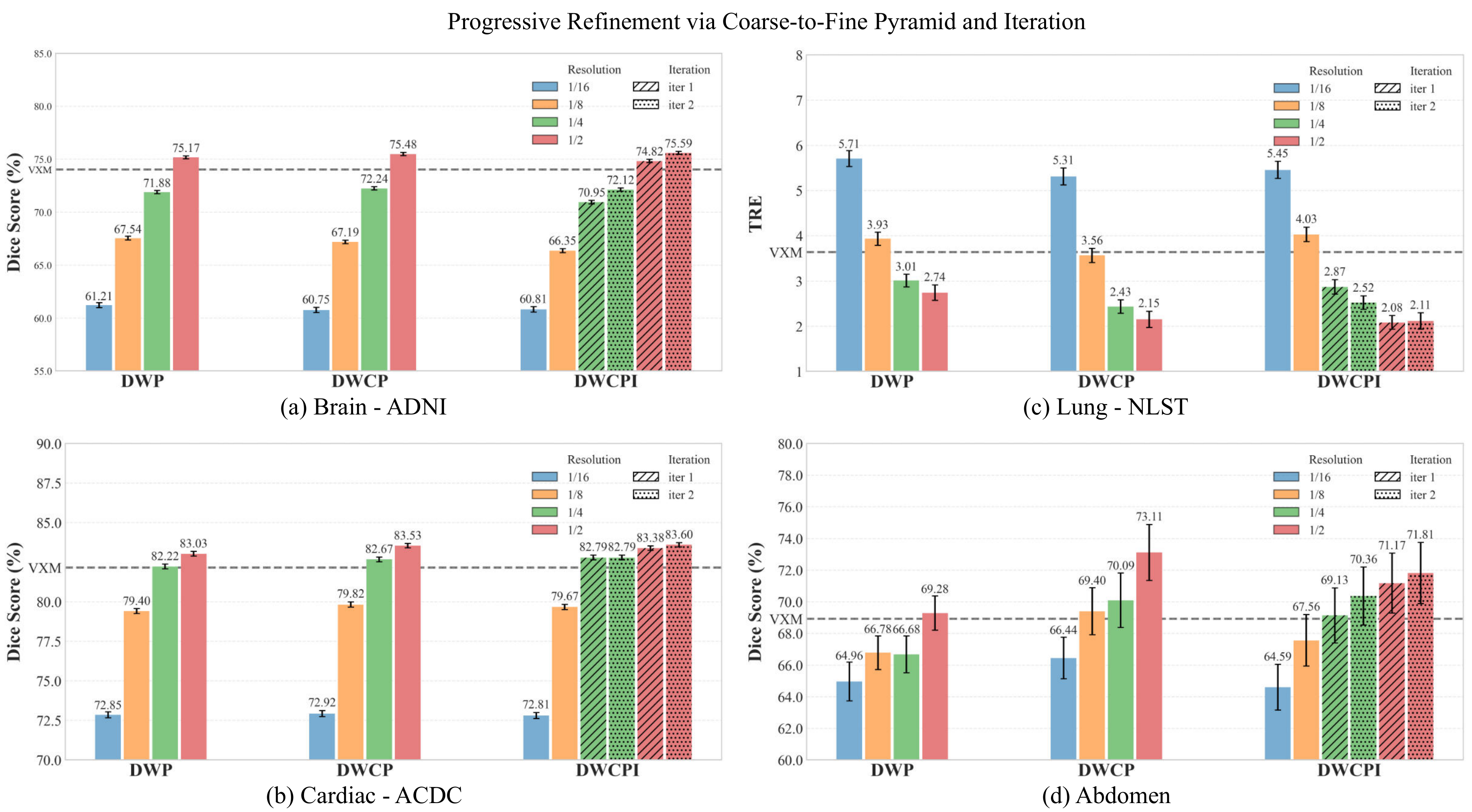}
    \caption{Progressive deformation refinement analysis across all tasks. Each subplot shows the incremental improvements in registration accuracy through coarse-to-fine pyramid refinement and iterative updates: (a) brain (ADNI), (b) cardiac (ACDC), (c) lung (NLST), and (d) abdomen anatomies. The VXM baseline reference is marked by the dotted horizontal line. Tables of additional metrics are provided in~\Cref{sec:prog_tabs}.
    }
    \label{fig:prog_bars}
\end{figure*}
\begin{figure}[ht!]
    \centering
    \includegraphics[width=\linewidth]{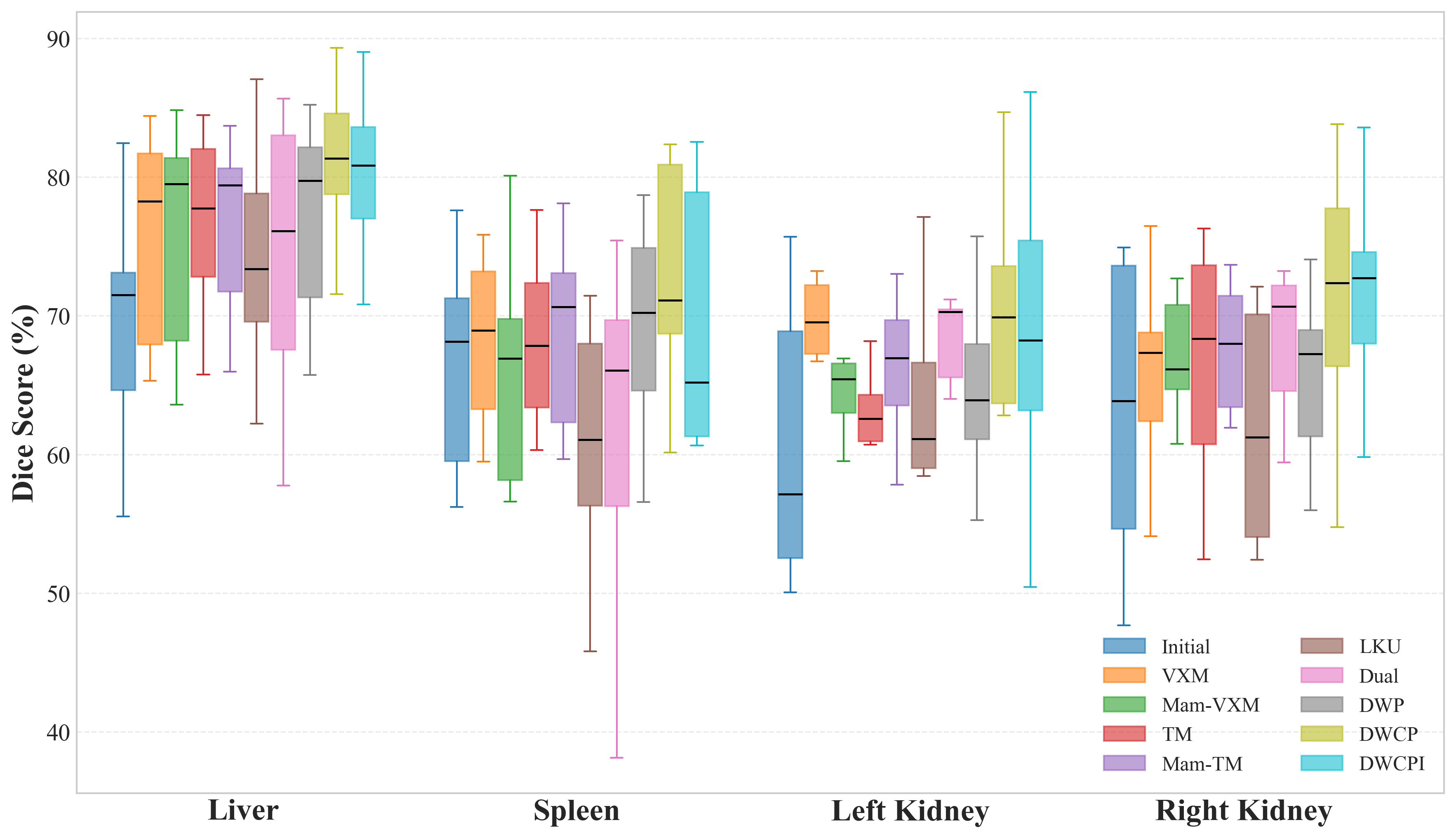}
    \caption{Box plots of structure-wise Dice scores on the Abdomen CT-MRI registration task. Structures include liver, spleen, left kidney and right kidney. A detailed breakdown of the Dice scores for each subject and organ is provided in~\Cref{sec:abd_appendix}.
    }
    \label{fig:organ_boxplot}
\end{figure}
\subsection{Intra-subject Abdomen MRI-CT Registration}
\begin{figure*}[ht]
    \centering
    \includegraphics[width=0.9\linewidth]{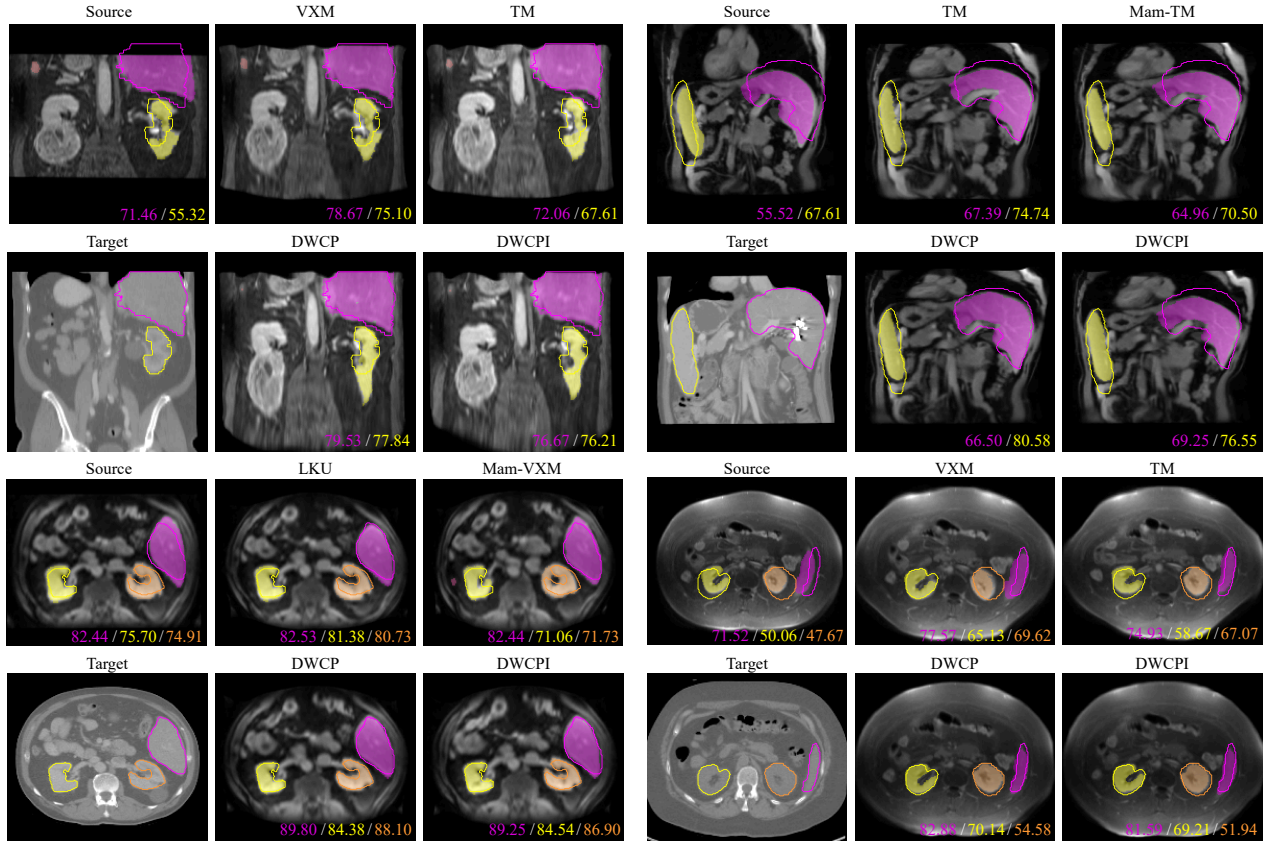}
    \caption{Qualitative registration results on Abdomen MRI-CT registration. The target image is shown with its segmentation contour overlaid. The source and registered images are overlaid with both their respective segmentation labels and the target contour.  Perfect registration aligns the contour precisely with the segmentation boundary. Dice scores for the liver, spleen, left/right kidney are reported within each image.
    }
    \label{fig:abd_qual}
\end{figure*}
\begin{table*}[ht!]
    \centering
    \footnotesize
    \caption{Quantitative results for the Abdomen MRI–CT registration task. Reported metrics include Dice score (DSC,\%), Normalized Surface Dice (NSD,\%), the standard deviation of the log-Jacobian determinant (SD$\log$J,$\times10^{-2}$), and the number of non-diffeomorphic voxels within the foreground (NDV,\textpertenthousand, per 10,000 voxels). The best (\colorboxlegend{C1}{6pt}, \textbf{bold}), second best (\colorboxlegend{C2}{6pt}, \textit{italic}), and third best (\colorboxlegend{C3}{6pt}) results are highlighted.}
    \label{tab:abdomen}
    \begin{tabular}{lcccccccc}
    \toprule
    & \multicolumn{4}{c}{MRI $\rightarrow$ CT} & \multicolumn{4}{c}{CT $\rightarrow$ MRI}\\
    \cmidrule(lr){2-5} \cmidrule(lr){6-9}
    & DSC $\uparrow$ & NSD $\uparrow$ & SD$\log$J $\downarrow$ & NDV(\textpertenthousand) $\downarrow$ & DSC $\uparrow$ & NSD $\uparrow$ & SD$\log$J $\downarrow$ & NDV(\textpertenthousand) $\downarrow$\\
    \midrule
    affine & 64.9$\pm$8.2 & 8.6$\pm$2.3 & - & - & 64.9$\pm$8.2 & 8.6$\pm$2.3 & - & -\\
    \midrule
    VXM & 69.0$\pm$3.9 & 11.1$\pm$1.8 & \cellcolor{C2}\textit{9.2$\pm$1.3} & \cellcolor{C3}0.2$\pm$0.3 & 68.9$\pm$4.5 & 12.5$\pm$1.2 & \cellcolor{C3}9.6$\pm$1.6 & \cellcolor{C3}3.6$\pm$4.1 \\
    Mam-VXM & 68.5$\pm$3.8 & \cellcolor{C3}11.1$\pm$1.7 & \cellcolor{C3}9.8$\pm$0.8 & 3.6$\pm$9.5 & 67.6$\pm$4.2 & 12.4$\pm$1.3 & 10.0$\pm$1.4 & 29.1$\pm$32.9 \\
    TM & 68.4$\pm$3.2 & 11.0$\pm$2.1 & 10.1$\pm$0.9 & 3.4$\pm$4.7 & 67.9$\pm$3.9 & 12.9$\pm$1.3 & 9.8$\pm$1.3 & 30.0$\pm$30.7 \\
    Mam-TM & \cellcolor{C3}69.6$\pm$3.5 & 11.0$\pm$1.9 & 10.2$\pm$1.0 & 21.9$\pm$56.6 & 67.5$\pm$4.1 & 12.5$\pm$1.5 & 9.8$\pm$1.2 & 41.8$\pm$52.1 \\
    LKU & 65.0$\pm$6.1 & 10.9$\pm$1.8 & \cellcolor{C1}\textbf{7.5$\pm$2.1} & \cellcolor{C1}\textbf{0.0$\pm$0.0} & 63.8$\pm$8.4 & 11.3$\pm$1.6 & \cellcolor{C1}\textbf{7.1$\pm$2.3} & \cellcolor{C2}\textit{0.1$\pm$0.2} \\
    \midrule
    Dual & 67.6$\pm$5.6 & \cellcolor{C2}\textit{11.2$\pm$1.2} & 10.1$\pm$1.0 & 2.9$\pm$3.5 & 67.3$\pm$5.4 & 12.5$\pm$1.2 & 10.0$\pm$1.3 & 15.0$\pm$21.3 \\
    DWP & 69.0$\pm$3.7 & 10.9$\pm$1.9 & 11.1$\pm$2.2 & 9.9$\pm$16.2 & \cellcolor{C3}69.6$\pm$4.6 & \cellcolor{C3}13.3$\pm$1.6 & 11.2$\pm$1.5 & 57.8$\pm$41.9 \\
    DWCP & \cellcolor{C1}\textbf{73.4$\pm$6.8} & \cellcolor{C1}\textbf{11.8$\pm$2.1} & 10.3$\pm$1.6 & 1.0$\pm$2.0 & \cellcolor{C1}\textbf{72.9$\pm$6.9} & \cellcolor{C1}\textbf{13.4$\pm$1.3} & 10.0$\pm$1.8 & 7.6$\pm$7.0 \\
    DWCPI & \cellcolor{C2}\textit{72.2$\pm$7.2} & 11.1$\pm$2.1 & 9.9$\pm$1.6 & \cellcolor{C1}\textbf{0.0$\pm$0.0} & \cellcolor{C2}\textit{71.5$\pm$7.7} & \cellcolor{C2}\textit{13.3$\pm$1.5} & \cellcolor{C2}\textit{9.3$\pm$1.3} & \cellcolor{C1}\textbf{0.0$\pm$0.0} \\
    \bottomrule
\end{tabular}
\end{table*}

\Cref{tab:abdomen} and~\Cref{fig:abd_qual} present both quantitative and qualitative results for intra-subject MRI–CT abdominal registration, one of the most challenging settings in our benchmark. Despite being intra-subject, the task involves substantial variations caused by changes in patient posture, field-of-view differences, and potential organ resection. Moreover, the contrast difference of the MRIs casts a significant domain shift for the multi-modal registration. These factors contribute to making this task a robust test of model generalization and robustness.

Consistent with other tasks, the low-level trend-driven architectures offer limited improvement over the VXM baseline. As shown in~\Cref{fig:organ_boxplot}, Transformer- and Mamba-based variants fail to significantly improve alignment accuracy across organs, suggesting that their general capacity for long-range modeling might not translate into effective cross-modal spatial matching. The LKU-Net model even underperforms the initial alignment on the spleen and right kidney, which could be attributed to the fact that the enlarged receptive field might not be helpful for the significant organ shifting.

In contrast, registration-specific designs, particularly the correlation volume combined with the motion pyramid (\textbf{DWCP}), yield the most substantial and consistent performance gains across all metrics (e.g., +4\% DSC, +0.9\% NSD). This demonstrates that explicitly computing modality-agnostic similarity via latent feature correlation is one of the keys to bridging the intensity gap between MRI and CT. Interestingly, the motion pyramid without the correlation volume (\textbf{DWP}) only achieves on-par performance with the VXM baseline and significantly degrades deformation smoothness.

Finally, the results for the iterative refinement (\textbf{DWCPI}) reveal a critical trade-off. While it successfully regularizes the deformation field, reaching lower SD$\log$J and near-zero folding voxels, this comes at the cost of a slight drop in registration accuracy compared to the single-shot \textbf{DWCP}. As illustrated in the top-right subject of~\Cref{fig:abd_qual}, the iterative refinement leads to poorer alignment of the kidneys, suggesting that the recurrent update may converge toward a suboptimal deformation.

\subsection{Progressive Deformation Refinement Analysis}
To better understand the cumulative benefits of our registration-specific components, namely, the motion pyramid, correlation volume, and iterative refinement, we perform a fine-grained analysis of registration accuracy across all refinement steps. For each variant, we evaluate metrics at every resolution level and iterative step to disentangle how these modules individually and collectively contribute to the final performance.

\paragraph{Motion pyramid achieves near-optimal alignment at surprisingly coarse scales} As shown in~\Cref{fig:prog_bars}, for most tasks and model variants, the registration accuracy at $1/4$ resolution already matches or even surpasses the VXM baseline (indicated by the VXM reference line crossing green bars). This demonstrates that a pyramidal design effectively captures large-scale deformations early, requiring only two or three coarse-to-fine refinements to achieve an anatomically meaningful alignment. In essence, most of the ``heavy lifting'' in deformation estimation happens at surprisingly low resolutions.

\paragraph{Correlation layer provides a strong inductive bias for voxel correspondence} When comparing \textbf{DWP} and \textbf{DWCP}, a clear and consistent trend emerges: adding the correlation layer boosts accuracy across all pyramid levels, with the largest gains observed at resolutions $1/4$ and $1/2$. This improvement highlights the importance of explicitly encoding local spatial similarity: correlation volumes act as an interpretable prior that reinforces voxel-to-voxel matching, complementing the implicit correspondence learned by the deformation decoder.

\paragraph{Iterative refinment improves fine details--but encourages ``lazy'' decoders} An interesting phenomenon arises when introducing iterative refinement (\textbf{DWCPI}). While iterations indeed improve the final accuracy at finer resolutions ($1/4$, $1/2$), as evidenced by the dashed versus dotted bars of the same color, they often lead to slightly reduced performance at coarser scales ($1/16$,$1/8$) compared to the single-shot \textbf{DWCP}. This suggests a ``lazy decoder'' behavior: once the network learns that an iterative update will follow, the early-stage deformation estimations relax their effort, deferring the correction to later iterations. This effect can be partially attributed to the gradient flow through the unrolled iteration, similar to recurrent networks, where accumulated gradients bias learning at a coarser level.
This insight opens an interesting avenue for future research: encouraging each resolution to reach near-optimal alignment before iterative refinement begins.

\paragraph{Summary}
Overall, this progressive refinement analysis provides concrete evidence that domain-specific architectural priors play a decisive role in improving deformable registration. The motion pyramid captures coarse global motion efficiently, the correlation layer enhances correspondence precision, and the iterative inference refines localized structures. Together, they form a principled hierarchy of complementary mechanisms, each contributing uniquely to robust and accurate deformation estimation across diverse imaging domains.

\section{Conclusion and Discussion}
\label{sec:conclu_discuss}
In this work, we revisited a fundamental question in learning-based deformable image registration: what truly drives progress?
Through a modular framework and comprehensive experiments across four diverse tasks, covering brain, lung, cardiac, and abdominal registration under various imaging modalities, we systematically disentangled the influence of low-level computation blocks from high-level, registration-specific designs.

Our findings consistently demonstrate that progress in registration is rooted in domain-specific inductive priors, rather than adopting ``more advanced'' or ``trend-driven'' architectures. While modern computational blocks such as Transformers, Mamba, or large-kernel convolutions offer marginal improvements, they come with substantial computational overhead and limited generalization. In contrast, registration-specific designs, including dual-stream encoder, motion pyramid, correlation, and iterative refinement, consistently yield more accurate, smoother, and more robust deformations across modalities and domains.

The correlation-based motion pyramid emerges as a particularly effective and generalizable design, bridging large displacement estimation and cross-modal matching. Meanwhile, the analysis of iterative refinement reveals both its promise and pitfalls, providing an intriguing research direction on balancing hierarchical and recursive optimization within a single model.

Beyond empirical findings, we release our modular, open-source benchmark as a foundation for fair, transparent, and extensible evaluation. This framework allows future studies to isolate the genuine contribution of new methods, ensuring that reported gains are not confounded by hidden priors or inconsistent training setups. We envision this as a living benchmark, continuously evolving with community contributions. (\url{https://github.com/BailiangJ/rethink-reg})

\paragraph{Limitations} While our study focuses on representative architectures and registration-specific designs, many paradigms imported from computer vision, such as diffusion-based registration, implicit neural representations (INR)-based deformations, or task-specific strategies like B-spline parameterization, remain unexplored. Nevertheless, these approaches can be seamlessly integrated into our modular benchmark. We view this work as an essential milestone toward a unified, extensible framework for fair and reproducible registration research. Besides, we do not carry out extensive hyperparameter, architectural, task-specific tuning to ensure fair and reproducible comparison across all methods. While this may not yield the absolute optimal performance for each method, it allows unbiased assessment of design principles and preserves the validity of our findings.

In summary, our results advocate a shift in research emphasis: from chasing architectural trends to understanding and leveraging domain priors. By grounding future work in principled, reproducible analysis, we hope to foster more transparent, generalizable, and clinically meaningful progress in medical image registration.

\paragraph{Acknowledgement}
This work was supported by BMWi (project ``NeuroTEMP'') research funding and the Munich Center of Machine Learning (MCML). We gratefully thank the computational resources provided by the Leibniz Supercomputing Centre (LRZ).
\paragraph{Disclosure of Interests}
The authors declare no competing interests.
\paragraph{Declaration of usage of LLMs}
During the preparation of this work, the authors used ChatGPT to improve its readability. After using this tool, the authors reviewed and edited the content as needed and take full responsibility for the content of the publication.

{
    \bibliographystyle{model2-names.bst}
    \biboptions{authoryear}
    \bibliography{main}
}
\clearpage
\appendix
\onecolumn
\section{Appendix}
\subsection{Progressive Refinement Analysis}\label{sec:prog_tabs}
\begin{table*}[ht]
    \caption{Progressive refinement analysis on  LPBA (40 subjects), and MindBoggle (100 subjects). OASIS (84 subjects), ADNI (43 subjects), and IXI (115 subjects) datasets, using 200 randomly sampled image pairs each. Reported metrics include Dice score (DSC, \%), and deformation regularity measures: SD$\log$J, ($\times 10^{-2}$). Metrics are reported across pyramid levels (1/16 $\rightarrow$ 1/2 resolution) and refinement iterations (Iter.~1, Iter.~2).}
    \label{tab:pyramid_brain}
    \centering
    \footnotesize
    \begin{tabular}{lcccccccccccc}
        \toprule
        & & & \multicolumn{2}{c}{LPBA} & \multicolumn{2}{c}{MindBoggle} & \multicolumn{2}{c}{OASIS} & \multicolumn{2}{c}{ADNI} & \multicolumn{2}{c}{IXI}\\
        \cmidrule(lr){4-5} \cmidrule(lr){6-7} \cmidrule(lr){8-9} \cmidrule(lr){10-11} \cmidrule(lr){12-13}
         & Res. & Iter. & DSC $\uparrow$ & SD$\log$J $\downarrow$ & DSC $\uparrow$ & SD$\log$J $\downarrow$ & DSC $\uparrow$ & SD$\log$J $\downarrow$ & DSC $\uparrow$ & SD$\log$J $\downarrow$ & DSC $\uparrow$ & SD$\log$J $\downarrow$\\
        \midrule
        \multicolumn{3}{l}{initial} & 54.0$\pm$14.2 & - & 51.5$\pm$24.5 & - & 57.3$\pm$22.4 & - & 52.8$\pm$22.1 & - & 54.5$\pm$23.7 & -\\
        \midrule
        \multirow{3}{*}{DWP} & 1/16 & - & 64.3$\pm$10.8 & 2.4$\pm$0.5 & 56.8$\pm$24.7 & 1.8$\pm$0.2 & 63.7$\pm$22.0 & 2.0$\pm$0.3 & 61.2$\pm$21.4 & 2.0$\pm$0.2 & 60.4$\pm$23.7 & 2.1$\pm$0.3 \\
        & 1/8 & - & 68.1$\pm$10.3 & 3.8$\pm$0.7 & 61.4$\pm$24.4 & 3.7$\pm$0.3 & 71.0$\pm$20.3 & 3.8$\pm$0.4 & 67.5$\pm$20.0 & 3.8$\pm$0.3 & 66.4$\pm$22.7 & 3.8$\pm$0.4 \\
        & 1/4 & - & 70.0$\pm$10.5 & 4.4$\pm$0.6 & 64.9$\pm$24.3 & 4.9$\pm$0.3 & 76.0$\pm$19.2 & 5.1$\pm$0.5 & 71.9$\pm$19.2 & 5.0$\pm$0.3 & 70.1$\pm$22.0 & 5.2$\pm$0.4 \\
        & 1/2 & - & 70.8$\pm$10.7 & 6.3$\pm$0.6 & 67.7$\pm$24.1 & 7.5$\pm$0.4 & 80.0$\pm$17.6 & 7.7$\pm$0.5 & 75.2$\pm$18.3 & 7.7$\pm$0.4 & 72.4$\pm$21.1 & 8.0$\pm$0.6 \\
        \midrule
        \multirow{3}{*}{DWCP} & 1/16 & - & 64.1$\pm$10.8 & 2.6$\pm$0.9 & 55.9$\pm$24.8 & 1.9$\pm$0.2 & 63.7$\pm$21.7 & 1.9$\pm$0.3 & 60.8$\pm$21.4 & 2.0$\pm$0.3 & 59.9$\pm$23.6 & 2.1$\pm$0.4 \\
        & 1/8 & - & 68.2$\pm$9.8 & 3.3$\pm$0.7 & 60.4$\pm$24.7 & 3.5$\pm$0.3 & 71.5$\pm$20.2 & 3.6$\pm$0.5 & 67.2$\pm$20.0 & 3.5$\pm$0.4 & 66.3$\pm$22.7 & 3.7$\pm$0.4 \\
        & 1/4 & - & 70.5$\pm$9.9 & 3.9$\pm$0.5 & 64.6$\pm$24.4 & 4.8$\pm$0.3 & 77.1$\pm$18.9 & 5.0$\pm$0.4 & 72.2$\pm$19.3 & 4.9$\pm$0.3 & 70.9$\pm$21.8 & 5.2$\pm$0.4 \\
        & 1/2 & - & 71.6$\pm$10.0 & 5.3$\pm$1.2 & 67.6$\pm$24.0 & 6.9$\pm$0.4 & 80.9$\pm$17.0 & 9.6$\pm$2.6 & 75.5$\pm$18.1 & 7.2$\pm$0.4 & 73.1$\pm$20.9 & 7.6$\pm$0.8 \\
        \midrule
        \multirow{5}{*}{DWCPI} & 1/16 & - & 64.0$\pm$10.8 & 2.6$\pm$0.8 & 55.5$\pm$24.8 & 1.8$\pm$0.2 & 63.3$\pm$22.0 & 1.8$\pm$0.3 & 60.8$\pm$21.1 & 2.0$\pm$0.3 & 59.9$\pm$23.7 & 2.0$\pm$0.4 \\
        & 1/8 & - & 67.1$\pm$10.1 & 3.3$\pm$0.7 & 58.9$\pm$24.9 & 3.5$\pm$0.4 & 70.3$\pm$20.8 & 3.5$\pm$0.5 & 66.3$\pm$20.4 & 3.5$\pm$0.4 & 65.2$\pm$23.1 & 3.5$\pm$0.4 \\
        & 1/4 & 1 & 69.4$\pm$10.1 & 3.5$\pm$0.6 & 63.2$\pm$24.6 & 4.0$\pm$0.3 & 75.7$\pm$19.4 & 4.1$\pm$0.5 & 70.9$\pm$19.5 & 4.1$\pm$0.3 & 69.7$\pm$22.2 & 4.2$\pm$0.4 \\
        & 1/4 & 2 & 70.0$\pm$10.1 & 4.0$\pm$0.6 & 64.2$\pm$24.4 & 5.0$\pm$0.3 & 77.0$\pm$19.0 & 5.0$\pm$0.5 & 72.1$\pm$19.2 & 5.0$\pm$0.3 & 70.6$\pm$21.9 & 5.2$\pm$0.4 \\
        & 1/2 & 1 & 71.2$\pm$10.3 & 4.7$\pm$0.6 & 66.8$\pm$24.1 & 6.1$\pm$0.3 & 80.3$\pm$17.4 & 6.7$\pm$0.9 & 74.8$\pm$18.4 & 6.2$\pm$0.3 & 72.7$\pm$21.1 & 6.4$\pm$0.4 \\
        & 1/2 & 2 & 71.3$\pm$10.3 & 5.6$\pm$0.6 & 67.7$\pm$23.9 & 7.2$\pm$0.4 & 81.1$\pm$16.8 & 10.2$\pm$3.1 & 75.6$\pm$18.0 & 7.4$\pm$0.4 & 73.2$\pm$20.7 & 7.7$\pm$0.5 \\
    \bottomrule
    \end{tabular}
\end{table*}

\begin{table*}[ht]
    \centering
    \footnotesize
    \caption{Progressive refinement analysis on NLST (in-domain) and Lung250M-4B (zero-shot) datasets. Reported metrics include target registration error (TRE, mm or vox), and deformation regularity measures: SD$\log$J, ($\times 10^{-2}$). Metrics are reported across pyramid levels (1/16 $\rightarrow$ 1/2 resolution) and refinement iterations (Iter.~1, Iter.~2).}
    \label{tab:lung_pyramid}
    \setlength{\tabcolsep}{4pt} 
    \begin{tabular}{lccccccccccc} 
        \toprule
        & & & \multicolumn{4}{c}{Lung250M-4B} & \multicolumn{4}{c}{NLST}\\
        \cmidrule(lr){4-7} \cmidrule(lr){8-11}
        & &
        & \multicolumn{2}{c}{INP $\rightarrow$ EXP} & \multicolumn{2}{c}{EXP $\rightarrow$ INP}  
        & \multicolumn{2}{c}{INP $\rightarrow$ EXP} & \multicolumn{2}{c}{EXP $\rightarrow$ INP}\\
        \cmidrule(lr){4-5} \cmidrule(lr){6-7} \cmidrule(lr){8-9} \cmidrule(lr){10-11}
        & Res. & Iter.& TRE $\downarrow$ & SD$\log$J $\downarrow$ & TRE $\downarrow$ & SD$\log$J $\downarrow$  & TRE $\downarrow$ & SD$\log$J $\downarrow$ & TRE $\downarrow$ & SD$\log$J $\downarrow$\\
        \midrule
        \multicolumn{3}{l}{initial} & 13.74$\pm$3.63 & - & 13.74$\pm$3.63 & - & 10.14$\pm$2.92 & - & 10.14$\pm$2.92 & - \\
        \midrule
        \multirow{3}{*}{DWP} & 1/16 & - & 11.49$\pm$3.42 & 5.7$\pm$1.7 & 10.82$\pm$3.39 & 5.2$\pm$1.4 & 5.77$\pm$1.72 & 4.1$\pm$1.0 & 5.64$\pm$1.80 & 4.1$\pm$1.0 \\
        & 1/8 & - & 10.16$\pm$3.84 & 8.0$\pm$1.7 & 9.89$\pm$3.78 & 6.8$\pm$1.4 & 3.96$\pm$1.46 & 5.3$\pm$1.1 & 3.91$\pm$1.47 & 5.5$\pm$1.2 \\
        & 1/4 & - & 9.78$\pm$4.14 & 10.5$\pm$2.2 & 9.53$\pm$4.06 & 8.7$\pm$1.7 & 3.01$\pm$1.41 & 6.6$\pm$1.3 & 3.00$\pm$1.40 & 6.7$\pm$1.4 \\
        & 1/2 & - & 9.64$\pm$4.22 & 13.1$\pm$2.7 & 9.51$\pm$4.16 & 10.9$\pm$2.1 & 2.60$\pm$1.44 & 7.9$\pm$1.5 & 2.60$\pm$1.41 & 7.9$\pm$1.5 \\
        \midrule
        \multirow{3}{*}{DWCP} & 1/16 & - & 10.38$\pm$3.11 & 6.2$\pm$1.1 & 9.92$\pm$3.27 & 4.7$\pm$0.6 & 5.34$\pm$1.90 & 4.0$\pm$0.9 & 5.28$\pm$1.81 & 4.1$\pm$0.9 \\
        & 1/8 & - & 8.94$\pm$3.46 & 8.6$\pm$1.9 & 8.34$\pm$3.57 & 6.4$\pm$1.0 & 3.57$\pm$1.55 & 5.5$\pm$1.1 & 3.56$\pm$1.55 & 5.6$\pm$1.2 \\
        & 1/4 & - & 8.26$\pm$3.86 & 10.6$\pm$2.2 & 7.77$\pm$3.83 & 7.8$\pm$1.1 & 2.43$\pm$1.47 & 6.6$\pm$1.2 & 2.44$\pm$1.47 & 6.7$\pm$1.3 \\
        & 1/2 & - & 8.04$\pm$4.02 & 12.7$\pm$3.1 & 7.58$\pm$3.91 & 9.5$\pm$1.7 & 2.00$\pm$1.45 & 7.5$\pm$1.2 & 2.03$\pm$1.46 & 7.6$\pm$1.5 \\
        \midrule
        \multirow{5}{*}{DWCPI} & 1/16 & - & 10.55$\pm$3.28 & 6.3$\pm$1.3 & 10.16$\pm$3.45 & 5.1$\pm$0.7 & 5.45$\pm$1.87 & 4.0$\pm$0.8 & 5.45$\pm$1.90 & 4.1$\pm$0.9 \\
        & 1/8 & - & 9.51$\pm$3.58 & 8.8$\pm$2.0 & 8.87$\pm$3.78 & 6.5$\pm$1.0 & 4.03$\pm$1.60 & 5.4$\pm$1.0 & 4.03$\pm$1.60 & 5.5$\pm$1.1 \\
        & 1/4 & 1 & 8.99$\pm$4.02 & 10.1$\pm$2.4 & 8.36$\pm$4.13 & 7.2$\pm$1.2 & 2.85$\pm$1.57 & 6.0$\pm$1.1 & 2.88$\pm$1.57 & 6.2$\pm$1.3 \\
        & 1/4 & 2 & 8.73$\pm$4.18 & 11.0$\pm$2.4 & 8.09$\pm$4.24 & 8.0$\pm$1.2 & 2.50$\pm$1.49 & 6.8$\pm$1.1 & 2.54$\pm$1.50 & 6.9$\pm$1.2 \\
        & 1/2 & 1 & 8.52$\pm$4.33 & 11.8$\pm$2.5 & 7.94$\pm$4.30 & 8.9$\pm$1.5 & 2.06$\pm$1.49 & 7.1$\pm$1.2 & 2.10$\pm$1.50 & 7.2$\pm$1.3 \\
        & 1/2 & 2 & 8.42$\pm$4.37 & 12.6$\pm$2.7 & 7.88$\pm$4.31 & 9.6$\pm$1.4 & 1.95$\pm$1.47& 7.6$\pm$1.1 & 2.00$\pm$1.49 & 7.7$\pm$1.3\\
        \bottomrule
    \end{tabular}
\end{table*}

\begin{table*}
\label{tab:cardiac_pyramid}
\footnotesize
    \centering
    \caption{Progressive refinement analysis on ACDC and M\&Ms (zero-shot) datasets. Reported metrics include Dice score (DSC, \%), normalized surface Dice (NSD, \%) and deformation regularity measures: SD$\log$J, ($\times 10^{-2}$). Metrics are reported across pyramid levels (1/16 $\rightarrow$ 1/2 resolution) and refinement iterations (Iter.~1, Iter.~2).}
    \setlength{\tabcolsep}{2.5pt} 
    \begin{tabular}{lccccccccccccccc} 
        \toprule
        & & & \multicolumn{6}{c}{M\&Ms} & \multicolumn{6}{c}{ACDC Test}\\
        \cmidrule(lr){4-9} \cmidrule(lr){10-15}
        & &
        & \multicolumn{3}{c}{ED $\rightarrow$ ES} & \multicolumn{3}{c}{ES $\rightarrow$ ED}  
        & \multicolumn{3}{c}{ED $\rightarrow$ ES} & \multicolumn{3}{c}{ES $\rightarrow$ ED}\\
        \cmidrule(lr){4-6} \cmidrule(lr){7-9} \cmidrule(lr){10-12} \cmidrule(lr){13-15}
        & Res. & Iter.
         & DSC $\uparrow$ & NSD $\uparrow$ & SD$\log$J $\downarrow$ 
         & DSC $\uparrow$ & NSD $\uparrow$ & SD$\log$J $\downarrow$ 
         & DSC $\uparrow$ & NSD $\uparrow$ & SD$\log$J $\downarrow$ 
         & DSC $\uparrow$ & NSD $\uparrow$ & SD$\log$J $\downarrow$\\
        \midrule
        \multicolumn{3}{l}{initial} & 60.5$\pm$12.7 & 8.0$\pm$6.0 & - & 60.5$\pm$12.7 & 8.0$\pm$6.0 & - & 60.2$\pm$14.8 & 10.7$\pm$7.3 & - & 60.2$\pm$14.8 & 10.7$\pm$7.3 & -\\
        \midrule
        \multirow{4}{*}{DWP} 
        & 1/16 & - & 72.7$\pm$12.0 & 12.6$\pm$5.1 & 2.0$\pm$0.7 & 74.3$\pm$11.3 & 10.3$\pm$5.1 & 1.8$\pm$0.6 & 69.5$\pm$14.3 & 15.2$\pm$5.8 & 2.3$\pm$0.8 & 70.7$\pm$13.6 & 11.8$\pm$5.5 & 1.8$\pm$0.6 \\
        & 1/8 & - & 77.6$\pm$10.6 & 15.3$\pm$5.6 & 2.5$\pm$0.8 & 81.1$\pm$9.2 & 14.5$\pm$6.7 & 2.3$\pm$0.7 & 76.1$\pm$12.7 & 18.3$\pm$6.1 & 3.4$\pm$1.2 & 79.5$\pm$11.2 & 16.4$\pm$6.4 & 3.1$\pm$1.0 \\
         & 1/4 & - & 79.8$\pm$9.7 & 17.9$\pm$6.0 & 3.0$\pm$1.0 & 84.0$\pm$8.1 & 16.9$\pm$7.2 & 2.7$\pm$0.8 & 79.3$\pm$11.5 & 21.9$\pm$6.0 & 4.3$\pm$1.5 & 83.5$\pm$9.5 & 20.4$\pm$6.9 & 3.7$\pm$1.2 \\
         & 1/2 & - & 80.3$\pm$9.5 & 18.8$\pm$6.3 & 3.5$\pm$1.2 & 85.3$\pm$7.6 & 18.8$\pm$7.8 & 3.2$\pm$1.2 & 80.3$\pm$10.9 & 23.3$\pm$5.7 & 4.9$\pm$1.6 & 85.4$\pm$8.5 & 23.2$\pm$7.5 & 4.3$\pm$1.3 \\
        \midrule
        \multirow{4}{*}{DWCP} 
        & 1/16 & - & 71.1$\pm$12.1 & 11.6$\pm$4.5 & 2.1$\pm$0.9 & 73.6$\pm$11.5 & 10.1$\pm$4.8 & 2.0$\pm$0.8 & 69.9$\pm$13.7 & 15.3$\pm$5.8 & 2.4$\pm$0.8 & 70.2$\pm$13.6 & 11.6$\pm$5.5 & 1.8$\pm$0.6 \\
        & 1/8 & - & 77.8$\pm$11.1 & 15.1$\pm$5.4 & 2.7$\pm$1.1 & 81.9$\pm$9.4 & 15.3$\pm$6.6 & 2.5$\pm$0.9 & 76.5$\pm$12.3 & 18.5$\pm$6.2 & 3.4$\pm$1.3 & 80.1$\pm$10.6 & 16.6$\pm$6.0 & 3.1$\pm$1.1 \\
         & 1/4 & - & 80.9$\pm$9.8 & 18.1$\pm$6.0 & 3.2$\pm$1.2 & 85.1$\pm$8.0 & 18.6$\pm$7.1 & 2.9$\pm$1.2 & 80.1$\pm$11.1 & 22.5$\pm$5.8 & 4.3$\pm$1.6 & 84.1$\pm$8.9 & 20.9$\pm$6.8 & 3.8$\pm$1.2 \\
         & 1/2 & - & 81.3$\pm$9.6 & 19.0$\pm$6.3 & 3.9$\pm$3.5 & 86.2$\pm$7.5 & 20.3$\pm$7.5 & 3.5$\pm$3.7 & 81.3$\pm$10.3 & 23.9$\pm$5.7 & 5.2$\pm$3.5 & 86.1$\pm$7.9 & 24.4$\pm$7.8 & 4.4$\pm$1.3 \\
        \midrule
        \multirow{6}{*}{DWCPI} 
        & 1/16 & - & 71.1$\pm$12.1 & 11.9$\pm$4.7 & 2.2$\pm$0.9 & 73.1$\pm$11.6 & 10.2$\pm$4.7 & 2.0$\pm$0.8 & 70.1$\pm$13.7 & 15.4$\pm$6.0 & 2.4$\pm$0.8 & 70.5$\pm$13.6 & 11.4$\pm$5.4 & 1.8$\pm$0.6\\
         & 1/8 & - & 77.9$\pm$11.1 & 15.2$\pm$5.6 & 2.7$\pm$1.1 & 81.5$\pm$9.5 & 15.3$\pm$6.5 & 2.5$\pm$0.9 & 76.2$\pm$12.4 & 18.2$\pm$6.4 & 3.4$\pm$1.3 & 79.5$\pm$11.1 & 16.4$\pm$6.3 & 3.1$\pm$1.0\\
         & 1/4 & 1 & 80.3$\pm$10.3 & 17.5$\pm$6.0 & 2.8$\pm$1.2 & 84.0$\pm$8.7 & 17.4$\pm$7.1 & 2.6$\pm$1.0 & 78.9$\pm$11.6 & 21.4$\pm$6.0 & 3.7$\pm$1.4 & 82.2$\pm$10.2 & 18.6$\pm$6.6 & 3.3$\pm$1.1 \\
         & 1/4 & 2 & 81.2$\pm$9.8 & 18.5$\pm$6.2 & 3.2$\pm$1.4 & 85.2$\pm$8.1 & 18.7$\pm$7.2 & 2.9$\pm$1.3 & 80.0$\pm$11.1 & 22.4$\pm$5.9 & 4.3$\pm$1.6 & 83.8$\pm$9.6 & 20.7$\pm$6.9 & 3.7$\pm$1.2 \\
         & 1/2 & 1 & 81.6$\pm$9.6 & 19.0$\pm$6.3 & 3.8$\pm$4.3 & 85.8$\pm$7.7 & 19.5$\pm$7.5 & 3.5$\pm$4.8 & 80.9$\pm$10.7 & 23.8$\pm$5.9 & 4.6$\pm$1.6 & 85.0$\pm$8.9 & 22.3$\pm$7.5 & 4.0$\pm$1.3 \\
         & 1/2 & 2 & 81.7$\pm$9.5 & 19.0$\pm$6.3 & 4.3$\pm$5.3 & 86.2$\pm$7.4 & 20.1$\pm$7.5 & 3.9$\pm$5.3 & 81.2$\pm$10.4 & 23.8$\pm$5.7 & 5.0$\pm$1.7 & 85.8$\pm$8.4 & 23.6$\pm$7.9 & 4.4$\pm$1.3 \\
        \bottomrule
    \end{tabular}
\end{table*}
\clearpage
\subsection{Abdomen MRI-CT DSC Breakdown}\label{sec:abd_appendix}
\begin{table*}[ht]
    \centering
    \caption{Dice score for each subject and organ on Abdomen MRI $\rightarrow$ CT registration task. The best (\colorboxlegend{C1}{6pt}, \textbf{bold}), second best (\colorboxlegend{C2}{6pt}, \textit{italic}), and third best (\colorboxlegend{C3}{6pt}) results are highlighted.}\label{tab:abd_organ_fwd}
    \vspace{5pt} 

    \footnotesize 
    \setlength{\tabcolsep}{4.0pt} 

    \begin{tabular}{l *{4}{cccc}}
    \toprule
    & \multicolumn{4}{c}{1} & \multicolumn{4}{c}{2} & \multicolumn{4}{c}{3} & \multicolumn{4}{c}{4} \\
    \cmidrule(lr){2-5} \cmidrule(lr){6-9} \cmidrule(lr){10-13} \cmidrule(lr){14-17}
    & Liver & Spleen & L-Kid & R-Kid & Liver & Spleen & L-Kid & R-Kid & Liver & Spleen & L-Kid & R-Kid & Liver & Spleen & L-Kid & R-Kid \\
    \midrule
    initial & 71.46 & 57.84 & - & 55.32 & 76.59 & \cellcolor{C2}\textit{71.63} & - & \cellcolor{C3}70.20 & 67.65 & \cellcolor{C2}\textit{71.04} & 55.40 & \cellcolor{C3}74.50 & 82.44 & 77.60 & 75.70 & \cellcolor{C3}74.91 \\
    \midrule
    VXM & 80.38 & 59.49 & - & 63.60 & 83.26 & 67.20 & - & 67.73 & \cellcolor{C1}\textbf{68.68} & \cellcolor{C3}70.71 & \cellcolor{C1}\textbf{66.72} & \cellcolor{C1}\textbf{76.48} & 84.41 & 64.51 & 70.27 & 66.92 \\
    Mam-VXM & 80.00 & 56.61 & - & 60.77 & \cellcolor{C1}\textbf{84.82} & 67.21 & - & 66.90 & \cellcolor{C3}67.72 & 58.65 & 59.50 & 72.68 & 83.40 & 66.62 & \cellcolor{C3}77.97 & 70.24 \\
    TM & 79.56 & \cellcolor{C3}60.32 & - & 64.28 & \cellcolor{C2}\textit{84.47} & 68.98 & - & \cellcolor{C1}\textbf{73.04} & 67.15 & 64.39 & 60.70 & \cellcolor{C2}\textit{76.29} & 83.31 & 66.70 & 68.17 & 62.00 \\
    Mam-TM & 78.80 & 59.66 & - & 61.93 & 83.70 & 70.30 & - & \cellcolor{C2}\textit{70.89} & 66.94 & 63.18 & 63.16 & 73.68 & 81.51 & 70.99 & 73.02 & 63.51 \\
    LKU & 78.65 & 56.39 & - & 54.34 & 79.33 & 58.46 & - & 62.78 & 62.22 & 45.80 & 59.69 & 69.91 & \cellcolor{C3}87.04 & 69.79 & 77.11 & 72.11 \\
    \midrule
    Dual & \cellcolor{C2}\textit{82.99} & 56.56 & - & \cellcolor{C3}69.76 & 83.03 & 65.50 & - & 66.27 & \cellcolor{C2}\textit{68.25} & 55.44 & \cellcolor{C3}64.01 & 71.57 & 85.67 & 70.03 & 70.52 & 71.85 \\
    DWP & \cellcolor{C1}\textbf{85.21} & 56.78 & - & 68.47 & 82.62 & \cellcolor{C1}\textbf{73.02} & - & 48.76 & 65.72 & 67.20 & \cellcolor{C2}\textit{66.27} & 70.48 & 82.00 & \cellcolor{C2}\textit{78.71} & 75.73 & 74.05 \\
    DWCP & 81.16 & \cellcolor{C1}\textbf{64.30} & - & \cellcolor{C1}\textbf{77.57} & \cellcolor{C3}84.41 & \cellcolor{C3}70.34 & - & 64.62 & 66.44 & \cellcolor{C1}\textbf{71.86} & 46.17 & 54.75 & \cellcolor{C1}\textbf{89.33} & \cellcolor{C1}\textbf{81.68} & \cellcolor{C2}\textit{84.67} & \cellcolor{C1}\textbf{83.82} \\
    DWCPI & \cellcolor{C3}81.26 & \cellcolor{C2}\textit{64.02} & - & \cellcolor{C2}\textit{76.73} & 80.43 & 60.63 & - & 59.80 & 64.10 & 60.79 & 50.44 & 57.16 & \cellcolor{C2}\textit{89.00} & \cellcolor{C3}77.96 & \cellcolor{C1}\textbf{86.13} & \cellcolor{C2}\textit{83.57} \\
    \end{tabular}

    \begin{tabular}{l *{4}{cccc}}
    \midrule
    & \multicolumn{4}{c}{5} & \multicolumn{4}{c}{6} & \multicolumn{4}{c}{7} & \multicolumn{4}{c}{8} \\
    \cmidrule(lr){2-5} \cmidrule(lr){6-9} \cmidrule(lr){10-13} \cmidrule(lr){14-17}
    & Liver & Spleen & L-Kid & R-Kid & Liver & Spleen & L-Kid & R-Kid & Liver & Spleen & L-Kid & R-Kid & Liver & Spleen & L-Kid & R-Kid \\
    \midrule
    initial & 55.52 & 56.22 & 51.56 & 52.58 & 71.52 & 60.08 & 50.06 & 47.67 & 45.42 & 65.26 & 58.88 & 57.53 & 71.93 & 71.15 & \cellcolor{C3}72.23 & \cellcolor{C2}\textit{73.30} \\
    \midrule
    VXM & 65.32 & 75.51 & 48.85 & 54.10 & 81.18 & 48.24 & \cellcolor{C2}\textit{72.85} & 58.82 & 65.64 & \cellcolor{C1}\textbf{72.43} & \cellcolor{C1}\textbf{73.21} & 67.90 & 76.13 & 75.85 & 68.82 & 71.43 \\
    Mam-VXM & 68.36 & 73.90 & \cellcolor{C3}62.22 & 65.36 & 80.69 & 40.01 & 65.57 & 72.39 & 63.58 & 68.39 & 65.33 & 65.24 & 79.02 & \cellcolor{C3}80.08 & 66.91 & 63.02 \\
    TM & \cellcolor{C2}\textit{74.69} & \cellcolor{C3}77.62 & 46.77 & 56.98 & \cellcolor{C3}81.62 & 48.95 & 61.62 & 72.38 & 65.75 & \cellcolor{C3}71.50 & 64.55 & \cellcolor{C1}\textbf{75.40} & 75.92 & 74.99 & 63.53 & 52.45 \\
    Mam-TM & \cellcolor{C3}73.32 & 75.11 & 57.82 & 63.08 & 80.32 & 45.23 & 64.66 & 68.29 & 65.95 & \cellcolor{C2}\textit{72.42} & \cellcolor{C3}69.22 & \cellcolor{C2}\textit{73.08} & \cellcolor{C3}80.05 & 78.09 & 69.85 & 67.68 \\
    LKU & 54.57 & 56.05 & 58.44 & 52.42 & 72.74 & \cellcolor{C1}\textbf{67.40} & 62.52 & 53.10 & \cellcolor{C2}\textit{73.98} & 63.68 & 58.77 & 59.72 & 72.02 & 71.44 & 67.98 & 70.66 \\
    \midrule
    Dual & 57.77 & 69.56 & 47.29 & 51.22 & 77.86 & 38.13 & 70.23 & \cellcolor{C3}73.23 & 65.39 & 66.62 & \cellcolor{C2}\textit{70.34} & 59.42 & 74.34 & 75.44 & 71.17 & \cellcolor{C3}73.23 \\
    DWP & 67.49 & 74.45 & 55.26 & \cellcolor{C3}67.25 & 80.30 & 56.58 & 68.51 & 67.24 & \cellcolor{C3}72.61 & 67.42 & 60.95 & 55.98 & 79.16 & 76.23 & 61.56 & 63.07 \\
    DWCP & \cellcolor{C1}\textbf{81.19} & \cellcolor{C2}\textit{82.35} & \cellcolor{C1}\textbf{66.30} & \cellcolor{C1}\textbf{78.31} & \cellcolor{C1}\textbf{85.11} & \cellcolor{C3}60.13 & \cellcolor{C1}\textbf{73.61} & \cellcolor{C2}\textit{73.58} & 71.57 & 70.19 & 62.82 & 66.93 & \cellcolor{C2}\textit{81.48} & \cellcolor{C2}\textit{80.63} & \cellcolor{C2}\textit{73.47} & 71.17 \\
    DWCPI & 70.80 & \cellcolor{C1}\textbf{82.52} & \cellcolor{C2}\textit{62.77} & \cellcolor{C2}\textit{71.95} & \cellcolor{C2}\textit{84.92} & \cellcolor{C2}\textit{61.47} & \cellcolor{C3}72.08 & \cellcolor{C1}\textbf{73.87} & \cellcolor{C1}\textbf{79.08} & 66.37 & 64.36 & \cellcolor{C3}70.73 & \cellcolor{C1}\textbf{83.16} & \cellcolor{C1}\textbf{81.76} & \cellcolor{C1}\textbf{76.54} & \cellcolor{C1}\textbf{73.47} \\
    \bottomrule
    \end{tabular}
\end{table*}

\begin{table*}
    \centering
    \caption{Dice score for each subject and organ on Abdomen CT $\rightarrow$ MRI registration task. The best (\colorboxlegend{C1}{6pt}, \textbf{bold}), second best (\colorboxlegend{C2}{6pt}, \textit{italic}), and third best (\colorboxlegend{C3}{6pt}) results are highlighted.}\label{tab:abd_organ_bck}
    \vspace{5pt} 

    \footnotesize 
    \setlength{\tabcolsep}{4.0pt} 

    \begin{tabular}{l *{4}{cccc}}
    \toprule
    & \multicolumn{4}{c}{1} & \multicolumn{4}{c}{2} & \multicolumn{4}{c}{3} & \multicolumn{4}{c}{4} \\
    \cmidrule(lr){2-5} \cmidrule(lr){6-9} \cmidrule(lr){10-13} \cmidrule(lr){14-17}
    & Liver & Spleen & L-Kid & R-Kid & Liver & Spleen & L-Kid & R-Kid & Liver & Spleen & L-Kid & R-Kid & Liver & Spleen & L-Kid & R-Kid \\
    \midrule
    initial & 71.46 & 57.84 & - & 55.32 & 76.59 & \cellcolor{C1}\textbf{71.63} & - & 70.20 & \cellcolor{C1}\textbf{67.65} & \cellcolor{C1}\textbf{71.04} & 55.40 & \cellcolor{C1}\textbf{74.50} & 82.44 & 77.60 & 75.70 & 74.91\\
    \midrule
    VXM & \cellcolor{C3}78.67 & 58.59 & - & \cellcolor{C3}75.10 & \cellcolor{C2}\textit{82.70} & 57.17 & - & \cellcolor{C3}75.51 & 65.58 & 59.51 & \cellcolor{C2}\textit{58.92} & 62.61 & 82.73 & 70.29 & 72.62 & 67.11\\
    Mam-VXM & 76.36 & 57.89 & - & 67.24 & 81.80 & 60.75 & - & 74.41 & 65.68 & 50.98 & 53.95 & 66.01 & 82.44 & 62.70 & 71.06 & 71.73\\
    TM & 72.06 & \cellcolor{C3}60.17 & - & 67.61 & 81.20 & 63.08 & - & \cellcolor{C2}\textit{76.89} & 63.55 & 62.37 & 57.70 & \cellcolor{C3}70.11 & 80.77 & 70.03 & 68.20 & 64.80\\
    Mam-TM & 74.37 & 58.12 & - & 66.88 & 81.30 & 59.21 & - & \cellcolor{C1}\textbf{79.29} & 64.89 & \cellcolor{C3}63.27 & 56.21 & 69.51 & 80.62 & 69.88 & 67.57 & 66.44\\
    LKU & 70.84 & 54.71 & - & 53.44 & 81.72 & 64.47 & - & 49.67 & 62.92 & 57.76 & \cellcolor{C3}57.80 & 67.35 & 82.53 & 70.28 & \cellcolor{C3}81.38 & \cellcolor{C3}80.73\\
    \midrule
    Dual & 77.80 & 53.61 & - & 72.00 & 82.29 & 58.98 & - & 74.25 & \cellcolor{C3}66.53 & 59.39 & 53.72 & 62.17 & \cellcolor{C3}83.16 & 71.68 & 64.44 & 73.44\\
    DWP & \cellcolor{C2}\textit{79.47} & 49.98 & - & 69.44 & \cellcolor{C1}\textbf{83.78} & \cellcolor{C3}64.64 & - & 70.21 & 66.21 & 62.42 & \cellcolor{C1}\textbf{59.37} & 69.88 & 82.26 & \cellcolor{C3}78.10 & 76.37 & 78.76\\
    DWCP & \cellcolor{C1}\textbf{79.53} & \cellcolor{C2}\textit{64.83} & - & \cellcolor{C1}\textbf{77.84} & \cellcolor{C3}82.32 & 64.47 & - & 74.44 & \cellcolor{C2}\textit{67.33} & \cellcolor{C2}\textit{69.08} & 43.88 & \cellcolor{C2}\textit{70.38} & \cellcolor{C1}\textbf{89.80} & \cellcolor{C2}\textit{82.30} & \cellcolor{C2}\textit{84.38} & \cellcolor{C1}\textbf{88.10}\\
    DWCPI & 76.67 & \cellcolor{C1}\textbf{65.20} & - & \cellcolor{C2}\textit{76.21} & 81.09 & \cellcolor{C2}\textit{67.84} & - & 68.49 & 62.26 & 59.92 & 52.16 & 67.45 & \cellcolor{C2}\textit{89.25} & \cellcolor{C1}\textbf{82.51} & \cellcolor{C1}\textbf{84.54} & \cellcolor{C2}\textit{86.90}\\
    \end{tabular}

    \begin{tabular}{l *{4}{cccc}}
    \midrule
    & \multicolumn{4}{c}{5} & \multicolumn{4}{c}{6} & \multicolumn{4}{c}{7} & \multicolumn{4}{c}{8} \\
    \cmidrule(lr){2-5} \cmidrule(lr){6-9} \cmidrule(lr){10-13} \cmidrule(lr){14-17}
    & Liver & Spleen & L-Kid & R-Kid & Liver & Spleen & L-Kid & R-Kid & Liver & Spleen & L-Kid & R-Kid & Liver & Spleen & L-Kid & R-Kid \\
    \midrule
    initial & 55.52 & 56.22 & 51.56 & 52.58 & 71.52 & \cellcolor{C1}\textbf{60.08} & 50.06 & 47.67 & 45.42 & 65.26 & 58.88 & 57.53 & 71.93 & 71.15 & \cellcolor{C3}72.23 & \cellcolor{C3}73.30\\
    \midrule
    VXM & \cellcolor{C3}66.92 & 71.03 & 54.17 & 62.47 & 77.57 & 44.32 & 65.13 & \cellcolor{C2}\textit{69.62} & 62.82 & 73.95 & \cellcolor{C2}\textit{78.50} & \cellcolor{C2}\textit{75.29} & 72.38 & 74.24 & 69.58 & \cellcolor{C2}\textit{76.00}\\
    Mam-VXM & 65.14 & 70.15 & \cellcolor{C3}61.78 & 62.60 & 78.18 & 48.60 & 68.17 & 64.78 & 60.77 & \cellcolor{C3}74.11 & 69.58 & 69.13 & 75.62 & 73.78 & 68.08 & 68.85\\
    TM & \cellcolor{C2}\textit{67.39} & \cellcolor{C3}74.74 & 57.60 & 60.96 & 74.93 & 46.43 & 58.67 & \cellcolor{C3}69.07 & 60.74 & \cellcolor{C1}\textbf{76.45} & \cellcolor{C3}70.79 & \cellcolor{C3}74.98 & 73.28 & 81.71 & 63.56 & 64.05\\
    Mam-TM & 64.96 & 70.50 & 52.37 & 62.80 & 72.12 & 47.92 & 61.20 & 67.17 & 63.60 & \cellcolor{C2}\textit{74.73} & 70.51 & \cellcolor{C1}\textbf{76.57} & \cellcolor{C3}76.62 & 81.17 & 60.59 & 59.14\\
    LKU & 58.53 & 52.75 & 53.93 & 52.71 & 68.14 & \cellcolor{C3}51.27 & 54.60 & 39.77 & 63.59 & 64.19 & 59.30 & 60.92 & 75.32 & \cellcolor{C2}\textit{83.82} & 70.30 & 70.37\\
    \midrule
    Dual & 60.77 & 67.76 & 47.71 & 52.25 & 77.54 & 40.43 & \cellcolor{C1}\textbf{73.68} & \cellcolor{C1}\textbf{74.81} & 61.64 & 69.67 & \cellcolor{C1}\textbf{78.89} & 65.57 & 71.06 & 74.11 & 66.36 & \cellcolor{C1}\textbf{78.05}\\
    DWP & 63.64 & 69.16 & 60.28 & \cellcolor{C3}67.98 & \cellcolor{C3}80.67 & \cellcolor{C2}\textit{51.31} & \cellcolor{C2}\textit{73.32} & 65.97 & \cellcolor{C3}67.48 & 71.17 & 69.65 & 64.06 & 75.97 & 77.86 & 66.68 & 70.98\\
    DWCP & 66.50 & \cellcolor{C1}\textbf{80.58} & \cellcolor{C1}\textbf{72.76} & \cellcolor{C1}\textbf{74.13} & \cellcolor{C1}\textbf{82.88} & 49.25 & \cellcolor{C3}70.14 & 54.58 & \cellcolor{C1}\textbf{73.46} & 72.73 & 66.63 & 72.93 & \cellcolor{C1}\textbf{83.53} & \cellcolor{C1}\textbf{84.15} & \cellcolor{C1}\textbf{73.01} & 67.78\\
    DWCPI & \cellcolor{C1}\textbf{69.25} & \cellcolor{C2}\textit{76.55} & \cellcolor{C2}\textit{64.21} & \cellcolor{C2}\textit{71.10} & \cellcolor{C2}\textit{81.59} & 42.09 & 69.21 & 51.94 & \cellcolor{C2}\textit{72.79} & 74.05 & 69.94 & 68.00 & \cellcolor{C2}\textit{82.37} & \cellcolor{C3}83.58 & \cellcolor{C2}\textit{72.42} & 72.58\\
    \bottomrule
    \end{tabular}
\end{table*}
\end{document}